\documentclass[natbib,namedreferences]{solarphysics}

\usepackage[optionalrh,solaenum]{solarphysics_submitted} 
\usepackage{graphicx}                    
\usepackage{subfig}
\usepackage{amssym b}                    
\usepackage{color}                       
\usepackage{url}                         

\graphicspath{{fig/}}

\begin{document}
\begin{article}
  \begin{opening}
    \title{3D Solar Null Point Reconnection MHD Simulations}
\author{G.~\surname{Baumann}$^{1}$\sep
        K.~\surname{Galsgaard}$^{1}$\sep
      \AA.~\surname{Nordlund}$^{1}$
   \\
   \sc \tiny Accepted to Solar Physics: October 10, 2012.
       }
\runningauthor{G. Baumann \textit{et al.}}
\runningtitle{Solar Null Point MHD simulations}
\institute{$^{1}$
   Niels Bohr Institute,
   Juliane Maries Vej 30,
   2100 K\o benhavn \O , Denmark
   email: \url{gbaumann@nbi.ku.dk}
}

\begin{abstract}
Numerical MHD simulations of 3D reconnection events in the solar corona
have improved enormously over the last few years, not only in
resolution, but also in their complexity, enabling more and more
realistic modeling. Various ways to obtain the initial magnetic
field, different forms of solar atmospheric models as well as diverse
driving speeds and patterns have been employed. This study
considers differences between simulations with stratified and
non-stratified solar atmospheres, addresses the influence of the
driving speed on the plasma flow and energetics, and provides
quantitative formulas for mapping electric fields and dissipation
levels obtained in numerical simulations to the corresponding solar
quantities.  The simulations start out
from a potential magnetic field containing a null-point, obtained from a
\textit{Solar and Heliospheric Observatory} (SOHO) magnetogram extrapolation approximately 8\,hours before a C-class flare
was observed. The magnetic field is stressed with a boundary motion
pattern similar to --- although simpler than --- horizontal motions observed
by SOHO during the period preceding the flare. The general behavior is
nearly independent of the driving speed, and is also very similar in stratified
and non-stratified models, provided only that the boundary motions are slow
enough. The boundary motions cause a build-up of current sheets, mainly
in the fan-plane of the magnetic null-point, but do not result in a
flare-like energy release. The additional free energy required
for the flare could have been partly present in non-potential form in the
initial state, with subsequent additions from magnetic flux emergence or
from components of the boundary motion that were not represented by the
idealized driving pattern.
\end{abstract}

  \end{opening}
\newcommand{\fig}[1]{Fig.\ \ref{fig:#1}} 
\newcommand{\Fig}[1]{Figure \ref{fig:#1}} 
\newcommand{\bea}[1]{\begin{eqnarray}\label{eq:#1}}
\newcommand{\eea}{\end{eqnarray}}
\newcommand{\beq}[1]{\begin{equation}\label{eq:#1}}
\newcommand{\eeq}{\end{equation}}
\newcommand{\Eq}[1]{Equation~\ref{eq:#1}}
\newcommand{\Eqs}[2]{Equations~\ref{eq:#1}--\ref{eq:#2}}
\newcommand{\Section}[1]{Section~\ref{sec:#1}}
\newcommand{\Table}[1]{Table~\ref{table:#1}}
\newcommand{\ddt}[1]{\frac{\partial #1}{\partial t}}
\newcommand{\ddx}[1]{\frac{\partial #1}{\partial x}}
\newcommand{\ddy}[1]{\frac{\partial #1}{\partial y}}
\newcommand{\ddz}[1]{\frac{\partial #1}{\partial z}}
\newcommand{\ddi}[1]{\frac{\partial #1}{\partial r_i}}
\newcommand{\ddj}[1]{\frac{\partial #1}{\partial r_j}}
\newcommand{\DDt}[1]{\frac{D #1}{ D t}}

\providecommand{\e}[1]{\ensuremath{\times 10^{#1}}}
\renewcommand{\div}{\nabla\cdot}
\newcommand{\curl}{\nabla\times}
\newcommand{\grad}[1]{{\nabla #1}}
\newcommand{\uu}{\mathbf{u}}
\renewcommand{\gg}{\mathbf{g}}
\newcommand{\pp}{\mathbf{p}}
\newcommand{\JJ}{\mathbf{J}}
\newcommand{\BB}{\mathbf{B}}
\newcommand{\EE}{\mathbf{E}}
\newcommand{\ux}{u_{\rm x}}
\newcommand{\uy}{u_{\rm y}}
\newcommand{\uz}{u_{\rm z}}
\newcommand{\gz}{g_{\rm z}}
\newcommand{\Hp}{H_{\rm P}}
\newcommand{\rr}{\mathbf{r}}
\newcommand{\Om}{\mathbf{\Omega}}
\newcommand{\Frad}{\mathbf{F}_{\rm rad}}
\newcommand{\Qrad}{Q_{\rm rad}}
\newcommand{\Qvisc}{Q_{\rm visc}}
\newcommand{\Qjoule}{Q_{\rm Joule}}
\newcommand{\Tvisc}{\mathbf{\tau}_{\rm visc}}
\newcommand{\Tij}{\tau_{ij}}
\newcommand{\sij}{s_{ij}}
\newcommand{\Bv}{B_{\nu}}
\newcommand{\Sv}{S_{\nu}}
\newcommand{\Iv}{I_{\nu}}
\newcommand{\tauv}{\tau_{\nu}}
\newcommand{\kv}{\kappa_{\nu}}
\newcommand{\Ekin}{E_{\rm kin}}
\newcommand{\Etherm}{E}
\newcommand{\Fconv}{\mathbf{F}_{\rm conv}}
\newcommand{\Fvisc}{\mathbf{F}_{\rm visc}}
\newcommand{\Fkin}{\mathbf{F}_{\rm kin}}
\newcommand{\half}{\frac{1}{2}}
\newcommand{\lbc}{{\rm bot}}
\newcommand{\Vpm}{\,V\,m$^{-1}$}

\section{Introduction}
\label{sec:introduction}
There have been different attempts to initialize the magnetic field of
the photosphere and corona for numerical simulations; amongst others by
elimination of the complex observed small scale structure
by the use of several photospheric magnetic monopole sources \citep{1997GApFD..84..127P}, by flux
emergence experiments \citep{2004A&A...426.1047A}, as well as by
extrapolation \citep[\textit{e.g.}][]{2009ApJ...700..559M} of solar
observatory magnetograms, \textit{e.g.} from SOHO.
The latter type has typically been used
together with potential extrapolations, for simplicity reasons as well as
due to the limited availability of vector magnetograms. As potential
magnetic fields contain no free magnetic energy, these cannot directly
be used for explaining how flare events take place and where the
released energy arises from. Therefore, to use a potential magnetic field
as the basis for an investigation of a flare event, the field must be
stressed into a state where it contains sufficient free magnetic energy
to account for the energy release event.
There are different ways by which this
may be accomplished. A simple approach is to impose boundary motions that
resemble the ones derived from observations \citep{2011A&A...530A.112B,2002ApJ...572L.113G}.
An alternative, more challenging approach is to stress
the system by allowing additional magnetic flux to enter through
photospheric magnetic flux emergence \citep{2003ApJ...589L.105F}. In the solar context both of
these processes take place simultaneously, while experiments typically
concentrate on a single type of stressing,
in order to investigate in detail its
influence on the dynamical evolution of the magnetic field.

The present investigation is an extension of the work done by
\cite{2009ApJ...700..559M}. They studied the evolution preceding
a specific flare event observed with SOHO, starting by taking a
magnetogram from
about 8 hours before the flare and deriving a potential magnetic
field. Due to the presence of a `parasitic' magnetic polarity,
the resulting magnetic field contains a magnetic null-point. From
the motions of observed magnetic fragments a schematic photospheric
velocity flow was constructed, and was used to stress this initially
potential magnetic field. The imposed stress distorts the
magnetic field, causing electric currents to build up in the vicinity of
the magnetic null-point. The magnetic dissipation associated with the electric
current allows a continuous reconnection to take place. The boundary driving together with the reconnection causes the null-point to move. The locations of the magnetic dissipation
agree qualitatively with the locations of flare emission in various
wavelength bands, which may be seen as evidence supporting a close
association between reconnection at the magnetic null-point and
the observed C-class flare.

The \citet{2009ApJ...700..559M} paper
raises many interesting questions, some of which we attempt to
answer in the present paper. We therefore let the same observations
provide the basis for deriving a potential initial magnetic field,
and employ the same imposed boundary stressing of the magnetic field, using
the setup to investigate the impact of varying the amplitude of the
driving speed, as well as the impact of allowing the experiment to take
place in a gravitationally stratified setting.

The paper is organized as follows:  In Section \ref{sec:methods} we
list the equations we solve, and briefly describe the numerical methods
used to solve them.  In Section \ref{sec:simulations} we give an
overview of the different numerical experiments, in Section
\ref{sec:results} we present and discuss the results, and finally in
Section \ref{sec:conclusions} we summarize the main results and
conclusions.

\section{Methods}
\label{sec:methods} The simulations have been performed using the fully 3D
resistive and compressible \emph{Stagger} MHD code \citep{Nordlund1997,2011ApJ...737...13K}.
The following form of the resistive MHD equations are solved in the
code:
\begin{eqnarray}
\frac{\partial \rho}{\partial t}
  &=& - \nabla \cdot (\rho \mathbf{u})
 \label{equ:continuity}\\
\frac{\partial (\rho \mathbf{u})}{\partial t}
  &=& - \nabla\cdot (\rho \mathbf{u} \mathbf{u} + \underline{\underline{\tau}})
  - \nabla p \nonumber \\
  &&
  + \mathbf{j} \times \mathbf{B}
  + \rho \mathbf{g}
 \label{equ:momentum}\\
\frac{\partial e}{\partial t}
  &=& - \nabla \cdot (e \mathbf{u} + \mathbf{f_e})
  - p \nabla \cdot \mathbf{u} \nonumber\\
  & &
  + Q_{J} + Q_{\nu}
 \label{equ:energy}\\
\frac{\partial \mathbf{B}}{\partial t}
  &=& - \nabla \times \mathbf{E}
 \label{equ:induction}\\
\mathbf{j} &=& \nabla \times \mathbf{B}
 \label{equ:current}\\
\mathbf{E} &=& - \mathbf{u} \times \mathbf{B} + \eta \mathbf{j}
 \label{equ:efield}\\
Q_J &=& \eta j^2
 \label{equ:Qmag}\\
\nabla \cdot \mathbf{B} &=& 0\\
p &=& (\gamma-1) e
 \label{equ:pressure}\\
\tau_{ij} &=&
  -\nu_{ij} \rho S_{ij}\\
S_{ij} &=&
  \frac{1}{2}\left(\frac{\partial u_i}{\partial x_j}
                 + \frac{\partial u_j}{\partial x_i} \right)\\
Q_\nu &=& \rho\sum_{ij}\nu_{ij} S_{ij}^2
  \label{equ:Qvisc}\\
\mathbf{f_e}
  &=& - \nu_e \rho \nabla (e/\rho)
\end{eqnarray}
where $\rho$ is the mass density, $\mathbf{u}$ the bulk velocity, $p$ the
pressure, $\mathbf{j}$ the current density, $\mathbf{B}$ the magnetic field,
$g$ the acceleration of gravity, $e$ the thermal energy per unit volume
and $\eta$ the resistivity. $S_{ij}$ is the shear tensor, $\tau_{ij}$
the viscous stress tensor and $\mathbf{f_e}$ is a weak diffusive flux
of thermal energy needed for numerical stability.
The term $Q_{v}$ represents viscous dissipation, turning
kinetic energy into heat, while $Q_{J}$ is the Joule dissipation,
responsible for converting magnetic energy into heat.

The solution to the MHD equations is advanced
in time using an explicit 3rd order predictor-corrector procedure
\citep{1979acmp.proc..313H}.

The version of the \emph{Stagger} MHD code used here assumes an ideal
gas law and includes no radiative cooling and heat conduction. The
variables are located on different staggered grids, which allows for
conservation of various quantities to machine precision.  The staggering
of variables has been chosen so $\nabla\cdot\BB$ is among the
quantities conserved to machine precision.
Interpolation of variables between different staggered grids is handled
by using 5th order interpolation. In a similar way spatial derivatives
are computed using expressions accurate to 6th order. To minimize the
influence of numerical diffusion dedicated operators are used for
calculating both viscosity and resistivity. The viscosity is given by
\begin{equation}\label{eq:nu}
    \nu = \Delta d \left(\nu_1 c_f + \nu_2 |\uu| +  \nu_3 \Delta d  \; |-\nabla\cdot\uu|_+ \right) ,
\end{equation}
where $\Delta d$ is the mesh size and
$\nu_1=0.005-0.02$, $\nu_2=0.005-0.02$,
and $\nu_3=0.2-0.4$ are dimensionless coefficients
that provide a suitable amount of dissipation of fast mode waves
($\nu_1$), advective motions ($\nu_2$), and shocks ($\nu_3$).
The expression $|-\nabla\cdot\uu|_+$ denotes the positive part of
the rate of compression $-\nabla\cdot\uu$. $c_f$ is the fast mode
speed defined by $c_f = \sqrt{(B^2 + \gamma p) / \rho} $.

The resulting grid Reynolds numbers, $\Delta d\, c_f/\nu$,
are on the order of 50 -- 200 in regions with smooth variations, while in the
neighborhood of shocks they are of the order of a few.
The corresponding expression for the resistivity is
\begin{equation}\label{eq:eta}
    \eta = \Delta d \left(\nu_1 c_f + \nu_2 |\uu| +  \nu_3 \Delta d \,|-\nabla\cdot\uu_{\perp}|_+ \right) ,
\end{equation}
where $\uu_{\perp}$ is the component of the velocity perpendicular
to $\BB$, and where the expression scaled by $\nu_3$ prevents
electric current sheets from becoming numerically unresolved.
The resulting magnetic grid Reynolds numbers are of the
order of a few in current sheets, as required to keep such structures
marginally resolved.

The overall scaling with $\Delta d$ ensures that
advection patterns, waves, shocks and current sheets
remain resolved by a few grids, independent of the mesh size.

The advantage of these three-part expressions for the viscosity and
the resistivity, compared to having constant viscosity and resistivity
is that constant values would have to be chosen on the order of the
largest of these three term, in order to handle shocks and current
sheets.  In the rest of the volume the viscosity and resistivity would
then be orders of magnitude larger than needed.  As demonstrated
in \citep{2011ApJ...737...13K} the results are quite similar to
state of the art codes that use local Riemann solvers. Such codes
also have dissipative behavior on the scale of individual cells --
no numerical code is 'ideal' in the sense that it presents solutions
corresponding to zero resistivity.

In this article we refer to the $-\uu \times \textbf{B}$ term in the induction equation as the advective electric field, while its counterpart $\eta \textbf{j}$ is referred to as the diffusive electric field.

\section{Simulations}
\label{sec:simulations}
The experimental setup is inspired by the work by
\citet{2009ApJ...700..559M}.
Our study sets out from a Fast Fourier Transform potential
extrapolation applied to a level 1.8 SOHO/\textit{Michelson
Doppler Imager} magnetogram \citep{1995SoPh..162..129S} from November 16, 2002 at
06:27\,UT, 8 hour prior to a C-class flare occurrence in the AR10191
active region.
The extrapolation leads to a 3D magnetic null-point topology with a
clear fan and spine structure \citep{Green89,1996RSPTA.354.2951P}.
In order to allow periodic boundary
conditions in the potential field extrapolation
we applied a windowing function to
the SOHO cutout of the active region AR10191, which decreases the field close to the boundary
towards zero. This cutout from the complete solar disk SOHO data differs
slightly from the one used by \citet{2009ApJ...700..559M}.
As a result, in our case the null-point is initially located at a height of about 4\,Mm above
the magnetogram, while in  their case the null-point is located at a height of
only 1.5\,Mm above the magnetogram. The change in null position
places our null-point in the corona proper and allows us to perform
affordable simulations with stratified atmospheres, while at the same
time it does not influence the nature of the current sheet formation, and
still successfully describes a typical solar-like magnetic field geometry.

Numerically, a magnetic field derived from a potential extrapolation
is not necessarily divergence free, and we therefore initially apply a
divergence cleaning procedure, which removes the divergence (as measured by our
specific numerical stretched mesh derivative operators) by iteratively applying a
correction obtained from solving the Poisson equation
\begin{equation}\label{eq:divb}
    \Delta \delta \Phi = -\nabla \cdot \mathbf{B} .
\end{equation}

\begin{figure}
    \centering
\subfloat[]{\includegraphics[scale=0.3]{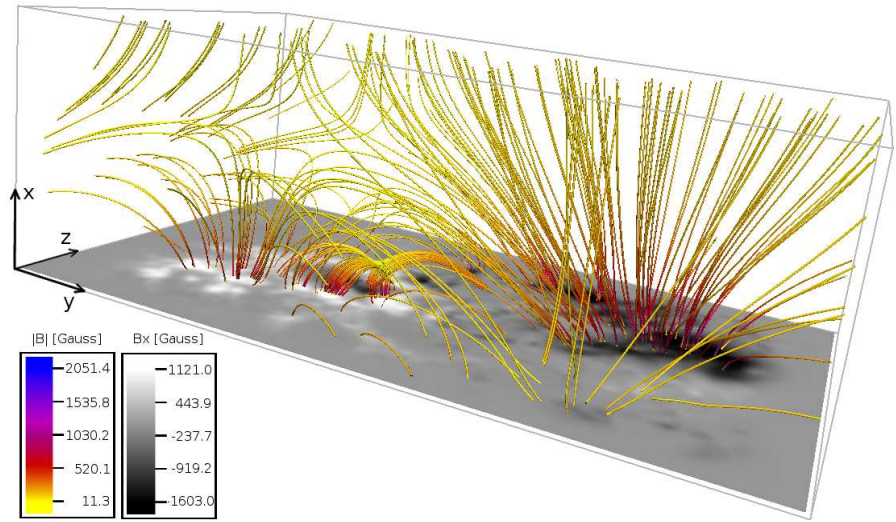}}\\
\subfloat[]{\includegraphics[scale=0.31]{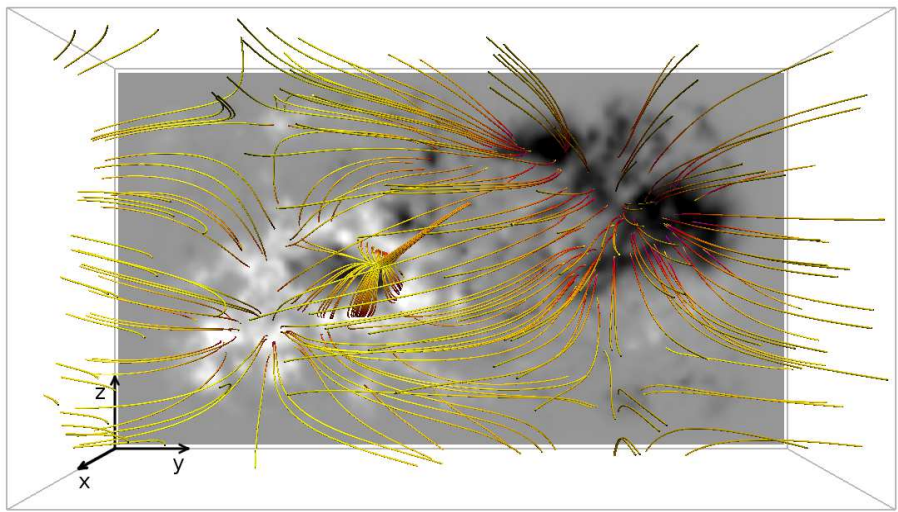}}
    \caption{Magnetic field resulting from the FFT extrapolation of
the SOHO magnetogram taken on November 16, 2002. The region shown here is
the entire computational box, having an extent of $60\times175\times100$\,Mm.
The slice represents the vertical component of the magnetic field.
Black is the negative polarity, white is positive.}
    \label{fig:init_MHD.eps}
\end{figure}

\begin{figure}
    \centering
        \includegraphics[width=0.5\linewidth]{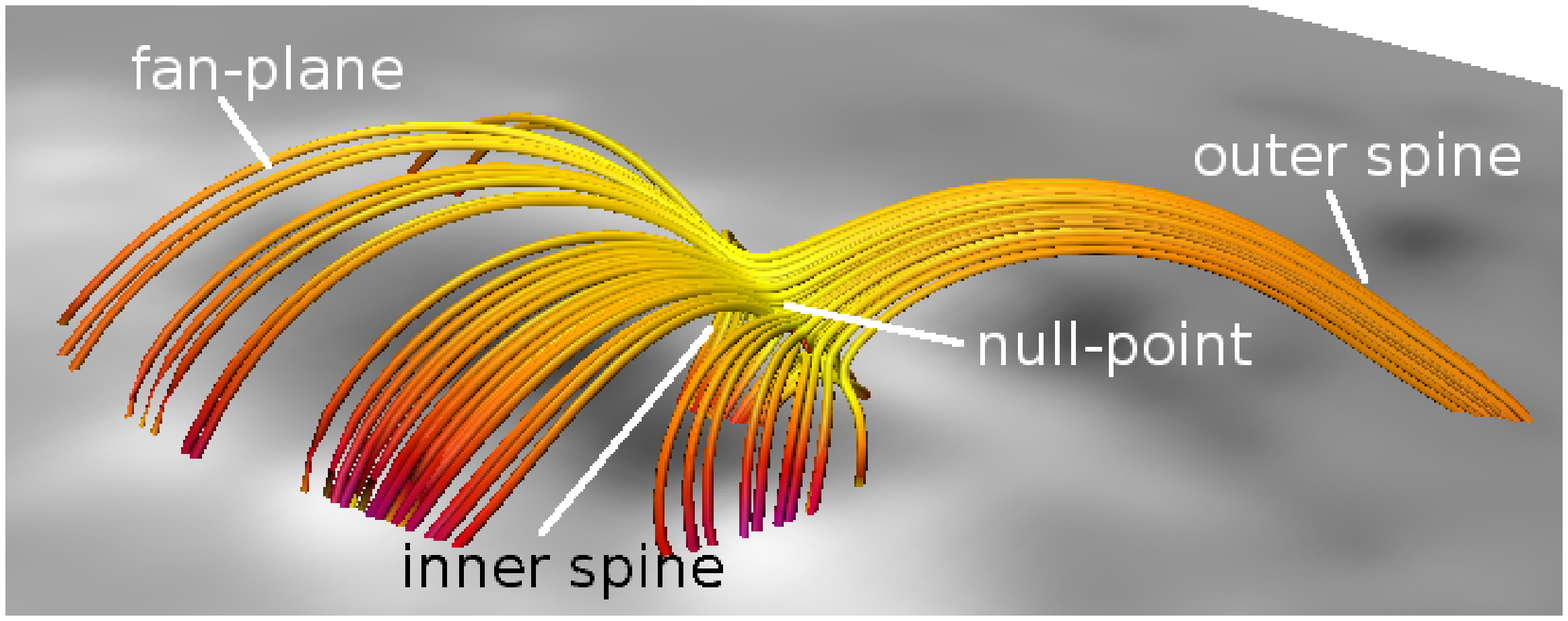}
    \caption{Zoom in of Figure \ref{fig:init_MHD.eps} (upper drawing) to the fan-spine
    topology, excluding the large scale field. The volume below the
    fan-plane is referred to as `the dome'.}
    \label{fig:spine_fan_zoom.eps}
\end{figure}

The structure of the resulting initial magnetic
field is illustrated in Figure \ref{fig:init_MHD.eps}a and
\ref{fig:init_MHD.eps}b, showing a strong overall magnetic field and a weaker
fan-spine structure close to the photospheric boundary, which is separately
illustrated in Figure \ref{fig:spine_fan_zoom.eps}. Note, that the field
lines of the fan-spine topology were selected specifically to show the topology,
and that the density of the field lines is therefore not representative of the
magnetic flux density.

\begin{figure}
    \centering
\includegraphics[scale=0.7]{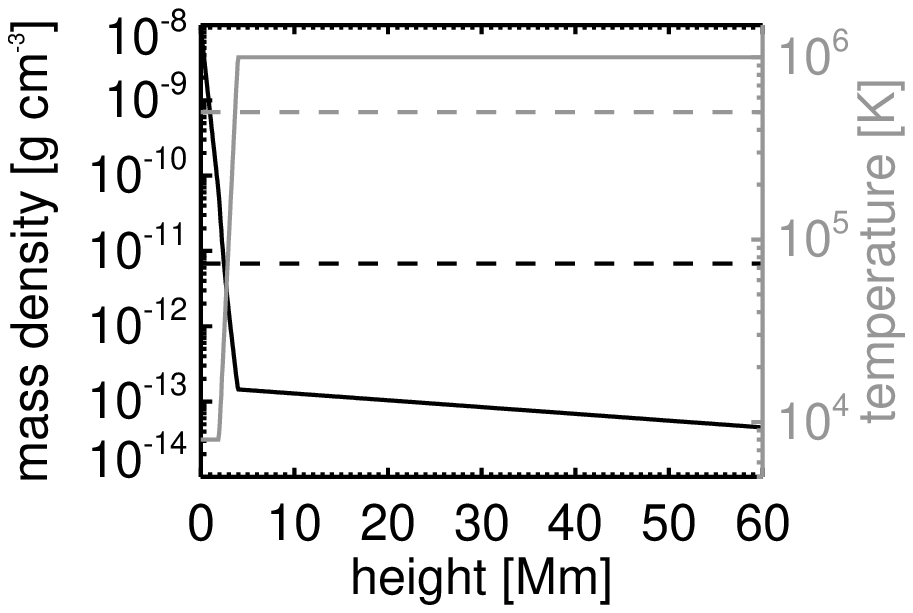}
    \caption{Mass densities are presented in black, temperatures in gray.
    The solid line shows the 1D mass density and temperature profiles as
    a function of height $x$ [Mm] of all stratified runs.
The dashed line
illustrates the constant mass density and temperature of run 1, run 1O, run 2
run 3 and run 4.}
    \label{fig:density_compare.ps}
\end{figure}

We performed two types of simulation; one type in which we imposed a
1-D gravitationally stratified atmosphere profile, and a second
type with constant
density and temperature. The density and temperature profiles are shown in Figure
\ref{fig:density_compare.ps}, and
a summary of the simulation runs may be found in Table
\ref{tab:simulations}.

The magnetic fields are in all cases anchored at the vertical boundaries, which
due to boundary conditions also prevent plasma from flowing in and out.
The exception to this is run 1O, in which instead constant pressure
is assumed at the lower boundary, and plasma flows through the boundary
are allowed. Only minor differences were found between the open and
closed boundary cases.

\begin{center}
   \begin{table*}
        \begin{tabular}{l c c c c c c c}
            \hline
 Run & boundary & max. driving [km\,s$^{-1}$] & density [cm$^{-3}$]& temperature [K]\\
       \hline
    1 & closed & 3.33 &6.8\e{12} &5.0\e{5}\\
    1O & open &  3.33 &6.8\e{12} &5.0\e{5}\\
    2 & closed & 6.67 &6.8\e{12} &5.0\e{5}\\
    3 & closed & 10   &6.8\e{12} &5.0\e{5}\\
    4& closed & 20   &6.8\e{12} &5.0\e{5}\\
    1S& closed & 3.33 &[4.5\e{10}, 9.1\e{15}] &[8000,1\e6]\\
    3S& closed & 10   &[4.5\e{10}, 9.1\e{15}] &[8000,1\e6]\\
    4S& closed & 20   &[4.5\e{10}, 9.1\e{15}] &[8000,1\e6]\\
       \hline
        \end{tabular}
    \caption{Simulation runs.
In the stratified runs min and max values are given in brackets.
`Boundary' refers to the plasma flow boundary condition at the
lower boundary. Runs with a stratified atmosphere are denoted
with an `S', while `O' stands for open boundary.}
    \label{tab:simulations}
   \end{table*}
\end{center}

An imposed horizontal velocity field is introduced at the lower boundary of the
computational box. This schematic velocity field
is based on the motions of magnetic fragments
observed by SOHO in the active region. These fragments outline the
fast relative motions observed prior to the flare in the active region
as discussed by \citet{2009ApJ...700..559M}.
We used and implemented the description of the velocity pattern provided in this reference.
In order not to produce initial transients
the velocity field at the bottom boundary is slowly ramped up,
over a period of about 100\,s, by using a hyperbolic tangent function,
and is afterwards kept constant.
A variety of driving speeds have been employed, ranging from values
similar to those used by \citet{2009ApJ...700..559M} to values about 6 times lower.
We compare and discuss their influence in Section \ref{sec:results}.

In general, it is important to keep the
driver velocity well below the Alfv\'en velocity of the magnetic
concentrations, because the magnetic structure and the plasma need to
have enough time to adapt to the changing positions of the magnetic
field lines at the boundary.  Ideally, to allow gas pressure to
equalize along magnetic fields, in response to compressions and
expansions imposed by the boundary motions, the driver velocity should
also be small compared to the sound speed in the coronal part of
the model.
This condition is generally fulfilled in all experiments,
since the coronal sound speed is on the order of 100\,km\,s$^{-1}$, while
our driving speeds are considerably smaller than that.

With these conditions we ensure an almost force free state at all times in the
simulation, which implies that the electric current is well aligned with the
magnetic field. Nevertheless, the line-tied motions of magnetic field lines
imposed by the lower boundary motions
causes the creation of a current sheet in which magnetic reconnection
takes place.

The maximal velocity of around 20\,km\,s$^{-1}$ that we applied exceeds the
actual velocities measured in the active region by a factor of about
40, while it is at the same time clearly sub-Alfv\'enic. This speed up,
which is similar to the one used by \citet{2009ApJ...700..559M},
has the desirable effect that we can cover a larger solar time
interval; a simulated time interval of 12 minutes then corresponds
to 8 hours of real solar time.  In our slowest cases (1S, 1, and 1O),
the simulated time is more than an hour, and the driving speed (3.33
km\,s$^{-1}$) is approaching realistic solar values.

\begin{figure}
    \centering
    \includegraphics[scale=0.5]{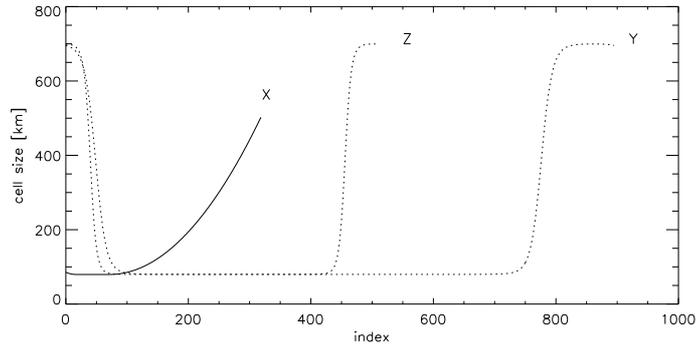}
    \caption{Cell sizes in km for all three axis plotted against the grid indices. X is the height. }
    \label{fig:grid_plot.ps}
\end{figure}

The simulated region has a size of $62\times175\times100$\,Mm, where
our $x$ axis points in the direction normal to the solar surface. The
computational box is covered by a
stretched grid of dimensions $320\times896\times512$, with a minimum
cell size of $\approx$ 80\,km maintained in a relatively large region
around the null-point. The grid size is smaller than 85 km over a
8\,x50\,x\,30\,Mm region, which includes the entire fan-plane and
its intersection with the lower boundary. The distributions of
cell sizes over grid indices are illustrated in Figure \ref{fig:grid_plot.ps}.
In the initial setup, the null-point is located at height index x = 50.

\section{Results and Discussions}
\label{sec:results}
As mentioned above, the field extrapolation based on the SOHO magnetogram
leads to a fan-spine topology of the magnetic field,
illustrated in Figure \ref{fig:spine_fan_zoom.eps}.
This
structure is surrounded by a stronger magnetic field, which
extends to much larger heights into the corona.  We
concentrate in the present study on the small fan-spine structure, which forms as a consequence of
a generally positive polarity in the active region AR10191 hosting a small
(`parasitic') negative polarity region. The overlying magnetic field lines,
including the ones
forming the fan-plane, are anchored in the photosphere and build
together with the spine a rather stable magnetic field structure,
keeping the plasma from expanding into the upper corona.

We simulate the motion of magnetic field lines located between
the large scale negative and positive polarities
--- hence outside the fan-spine structure --- which on November 16, 2002
moved a large amount of magnetic flux towards the east side\footnote{We
use  as reference system the solar coordinate system.} (left hand
side in Figure \ref{fig:init_MHD.eps}) of the dome, which we define
as the volume confined by the fan-plane. This translational motion
at the photospheric
boundary is represented in our experiment by a boundary motion (`driver'),
which is applied at the lower boundary of our computational box.
The boundary motions lead to an eastward
directed motion of the magnetic field lines outside the dome and to magnetic plasma being pushed
against the west periphery of the fan-spine structure.
Especially in the stratified case, a part of this flow extends
upward along the magnetic field lines toward the neighborhood of the
null-point and the outer spine.

\begin{figure}
    \centering
        \includegraphics[scale=0.195]{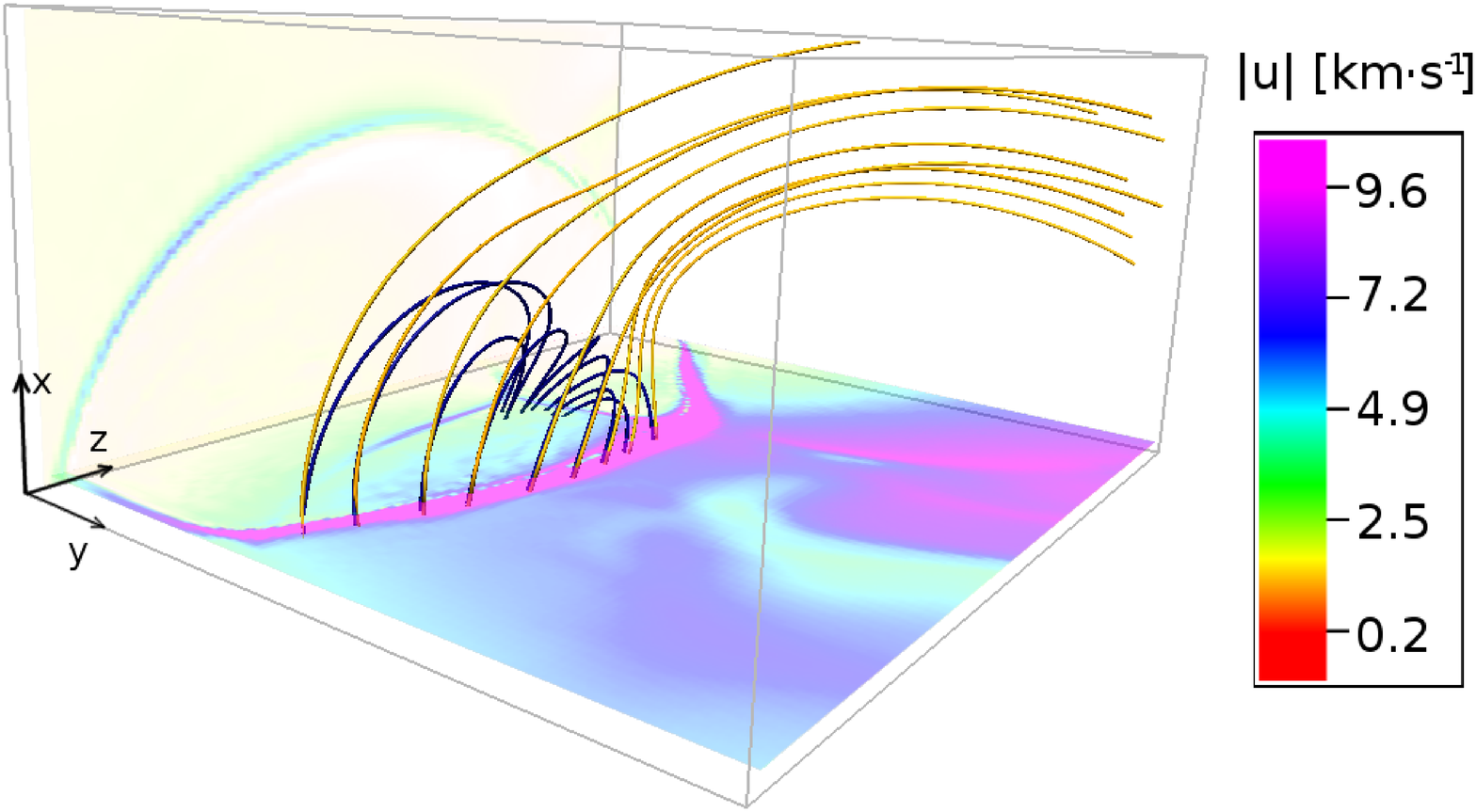}
        \includegraphics[scale=0.245]{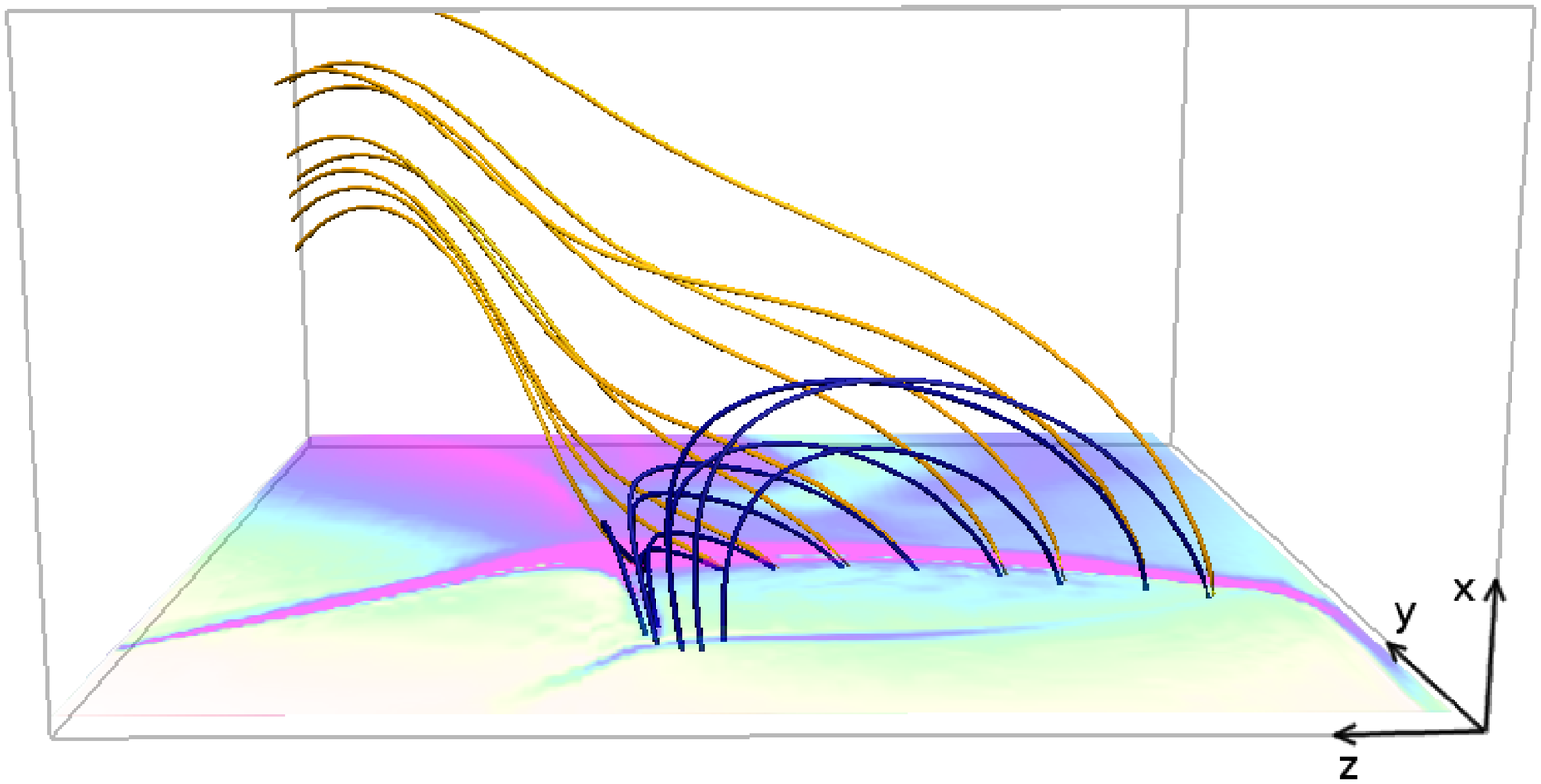}
    \caption{Magnetic field lines connecting to the inside (dark blue)
 and to the outside (orange) of the fan-plane. The bottom slice shows the
bulk speed at a height of about 1.4\,Mm.  The snapshot is taken at a displacement D = 2.15\,Mm
in the experiment run 3. The box size is $10\times16\times22$\,Mm.}
    \label{fig:fieldline_shear}
\end{figure}

The displacement of field lines, particularly outside of the fan-plane,
introduces a misalignment between the inner and outer spine (see also
Figure \ref{fig:spine_fan_zoom.eps}).
Figure \ref{fig:fieldline_shear} shows the field line shear
at a nominal `boundary displacement' (simulation time times the amplitude
of the average applied boundary velocity),
D = 2.15\,Mm after the start of run 3, which has a driving
speed of about 10\,km\,s$^{-1}$. Choosing the displacement instead of the simulation time has the advantage that at
a given displacement, all runs have experienced about the same energy input
from the work introduced by the boundary driving and are hence comparable. The angle $\phi$, designating
the difference of the direction between the inner magnetic field lines (in dark blue)
and the outer magnetic field lines (in orange) with respect to the fan-plane, is still
quite small. The quasi-transparent slices show the bulk
speed, which is high just outside the fan-plane,
where plasma is pushed up by the driver. Magnetic field lines closely
approaching the null-point run just below this high bulk flow
layer.

\begin{figure}
    \centering
        \includegraphics[scale=0.17]{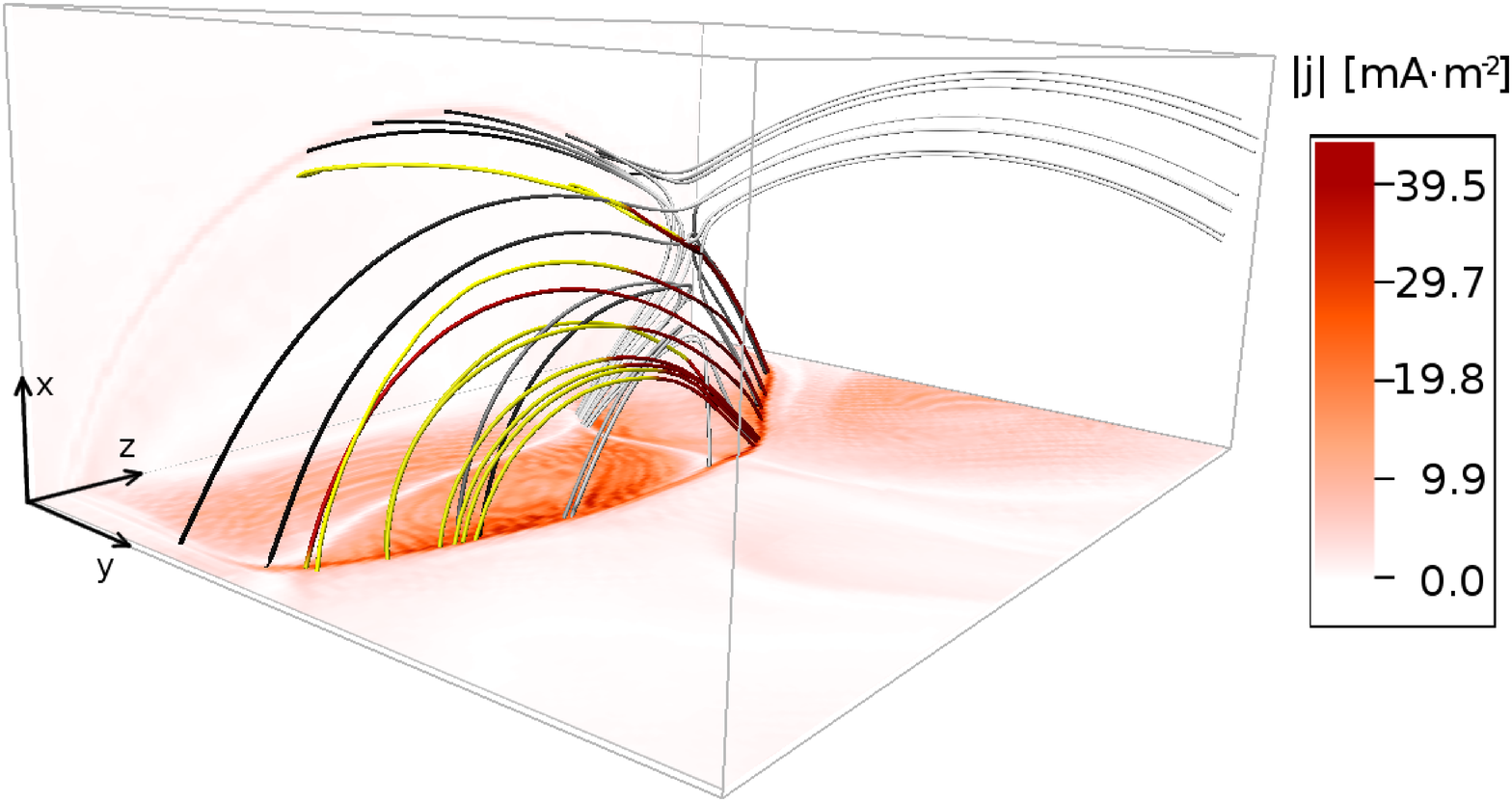}
        \includegraphics[scale=0.18]{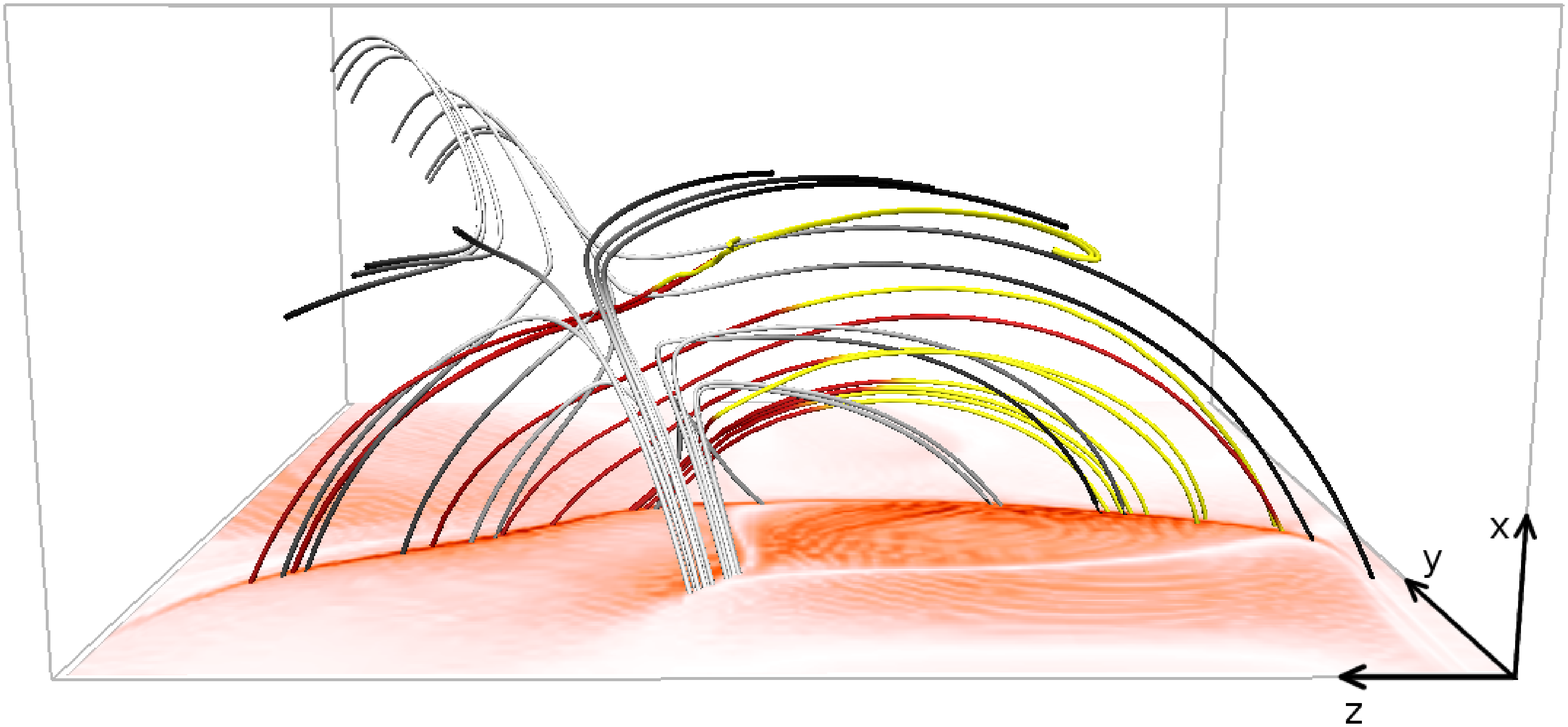}
    \caption{The electric current (red--yellow) flowing antiparallel
to the magnetic field lines (black--gray). The
color gradient along the streamlines indicates the flow direction (going from
red to yellow) and the magnetic field orientation (going from black to gray).
The bottom slices show the electric current density of run 3. D =
2.68\,Mm and the box size is $10\times16\times22$\,Mm.}
    \label{fig:j_fanplane}
\end{figure}

The applied photospheric driving motion indirectly moves the fan-plane foot
points at the west side of the fan-spine structure, causing a
slight shear between the inner and the outer field
lines of the fan-plane to arise due to a different stress level of these
two flux systems. The magnetic flux system inside the dome experiences
a compression of about 5 times the surrounding gas pressure
when the magnetic flux system west to the dome has moved towards it. This
leads to a large stress close to the null-point, on the east side of
the outer spine, where the magnetic field as a consequence reconnects with the surrounding field in
order to reach a lower energy state. A thin current sheet forms in the
fan-plane, with the largest electric current densities occurring
closest to the driver, where the shear of the field lines is
largest. The magnitude of the electric current is lower in the neighborhood of the
null-point. At the null-point itself the effect is to disrupt the structure
of the null in such a way that the two spine axes move apart, as
seen also in other single null investigations
\citep{2007PhPl...14e2106P,2011A&A...529A..20G}.  Figure
\ref{fig:j_fanplane} shows the streamlines and direction of the highest electric current in
run 3 at D = 2.68\,Mm. We find that the electric current is mostly
anti-parallel to the magnetic field lines in the fan-plane.

A partial outcome of the reconnection is seen in the motion of the inner
spine, which is not being moved directly by the applied photospheric
driver, but nevertheless moves at the photospheric level a significant absolute distance of about
4.5\,Mm in the simulation.
The motion is nearly linear in space and time.
A second signature is the character of the displacement of the outer spine
relative to the position of the inner spine.

\begin{figure}
    \centering  \includegraphics[scale=0.17]{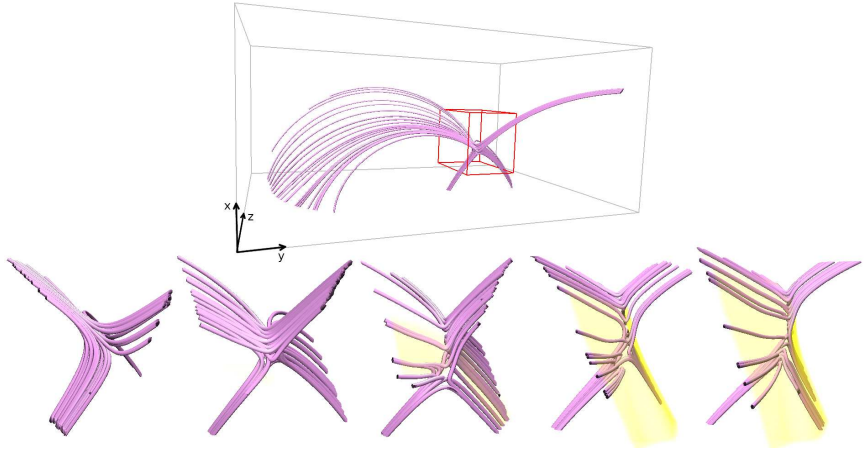}
    \caption{Relative motion of inner and outer spine for run 3. The
top panel shows the cutout region for D = 0.03\,Mm and the lower
panel shows approximately the same selected region for
D = 0.03, 0.55, 1.08, 1.62, 2.15\,Mm. The yellow volume indicates the
highest electric current
density, which is located in the fan-plane.}
    \label{fig:bfield}
\end{figure}

The initial null-point area, connecting the inner and outer spine, stretches with
increasing displacement into a `weak field region', where the magnetic field
strength is very low. This initiates an electric current that passes through
the fan-plane, causing the ratio of the smallest to largest
fan-eigenvalues to decrease \citep{1996PhPl....3..759P}. In this case the null
almost adopts a 2D structure. This is illustrated in Figure \ref{fig:bfield}. We discuss
the relation of this inner and outer spine distance to the electric field
at the end of section \ref{sec:efield}.

\subsection{Time evolution of the diffusive electric field\label{sec:efield}}
\begin{figure*}[tbp]
    \centering
    \includegraphics[scale=0.6]{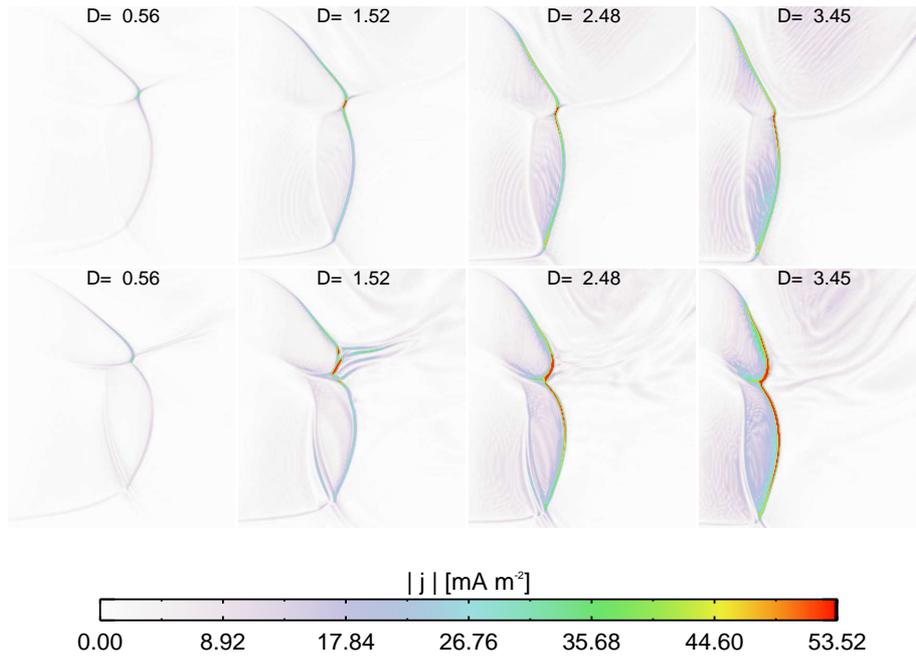}
    \caption{Current density in y-z slice for constant density run 4 (upper row) and
stratified atmosphere run 4S
(lower row) at a height of x = 3.22\,Mm and different displacements D.}
    \label{fig:j_condense_strati_compare.ps}
\end{figure*}

The diffusive part of the electric field ($\eta \textbf{j}$) included in the induction equation is responsible for
both changing the magnetic field topology and for transforming magnetic
energy into Joule dissipation in the MHD picture. In the Sun the
diffusive electric field component parallel to the magnetic field is responsible for particle
acceleration \citep{2006A&A...454..957A}. It is therefore of particular
interest to see how this field evolves with the boundary displacement,
and to find out where it concentrates and how large values it
reaches. Figure \ref{fig:j_condense_strati_compare.ps} illustrates
the electric current accumulation in the fan-plane and along the spine axes
of the magnetic null. This behavior is representative for all stratified and
non-stratified runs. In the code the resistivity is not
a simple constant, as specified by Equation (\ref{eq:nu}),
allowing diffusion
to be locally increased where dissipation is needed to keep structures
from becoming unresolved, while at the same time allowing a minimal amount
of diffusion in regions where the magnetic field is smooth. Images of the diffusive
electric field are therefore not exact replicas of images of the electric
current, but since the resistivity is generally near its largest value in
current sheets there is a close correspondence.

The diffusive electric
field is found to be concentrated in the fan-plane and to have its local
peak in the region where the fan-spine intersection is distorted. However,
large values occur over a significant fraction of the fan-plane, as is
the case for the electric current density.

\begin{figure}
    \centering
\subfloat[]{\label{fig:electric_field_a}
\includegraphics[scale=0.46]{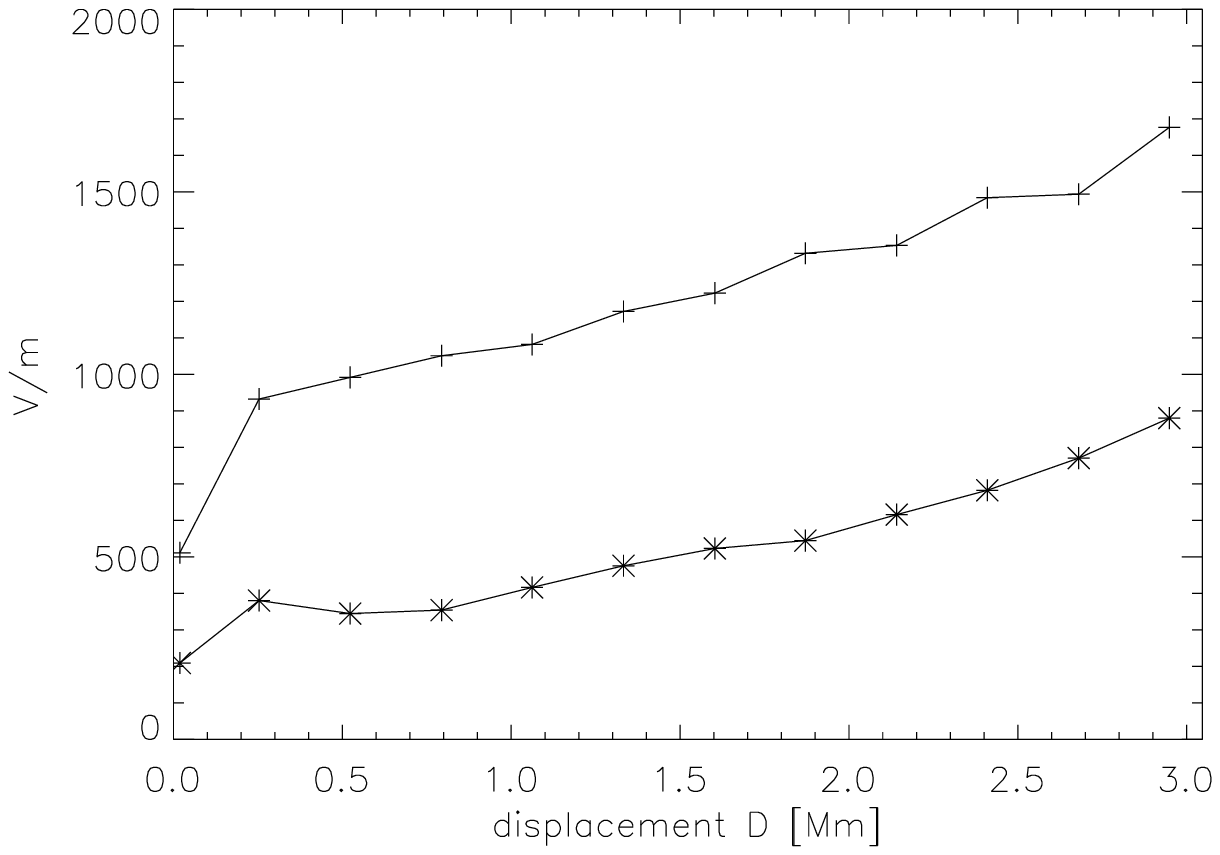}}\hspace{-0.5 cm}
\subfloat[]{\label{fig:electric_field_b} \includegraphics[scale=0.46]{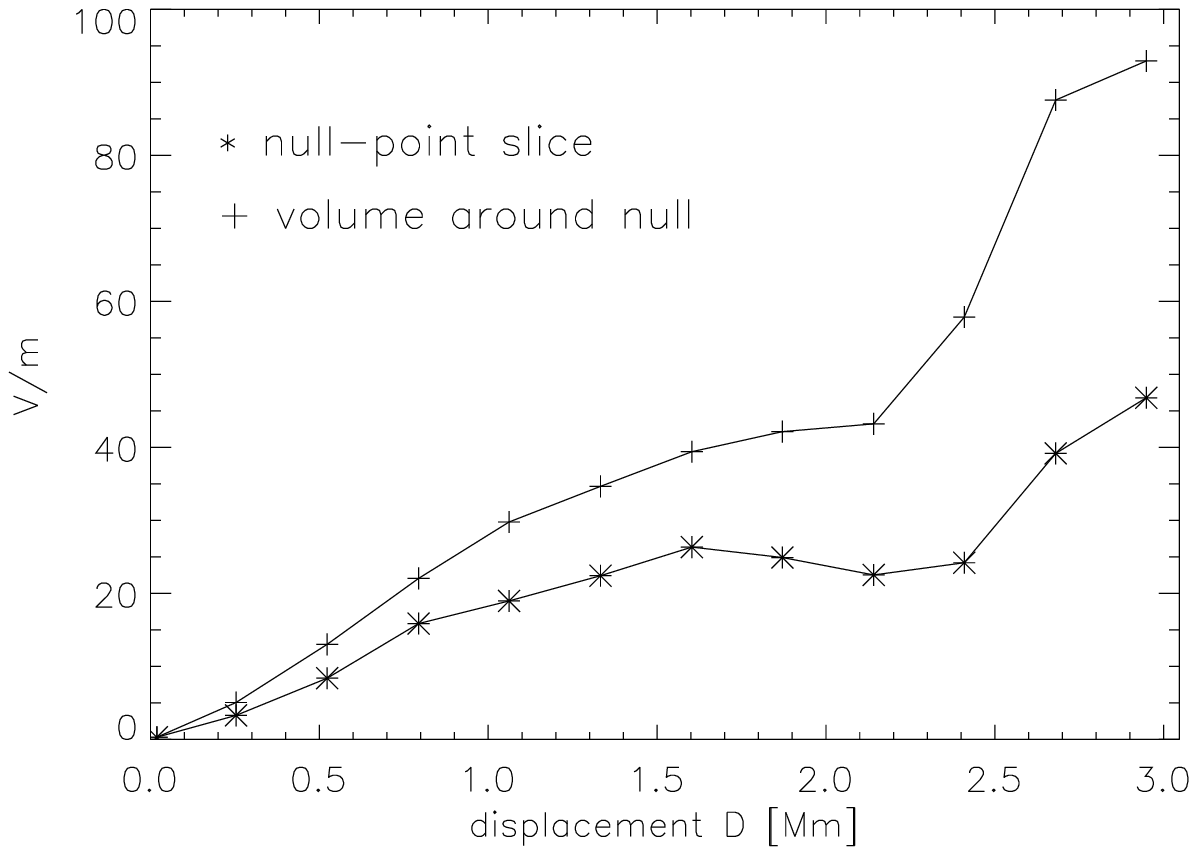}}
    \caption{Maximum advective electric field (a) and average diffusive electric
    field calculated over those grid points at which the electric field amplitude
    is within 10\,\% of the maximum field in a slice of 2 grid cells thickness
    (b) of the cutout region at the respective height of the null-point (*) and
    in the total cutout region $10\times16\times22$\,Mm around the null-point,
    including the fan-plane (+) in run 3.}
    \label{fig:electric_field}
\end{figure}

It is only the parallel
diffusive electric field that gives rise to both magnetic
reconnection and particle acceleration and its magnitude indicates how violent these processes can be \citep{1988JGR....93.5547S}. In our simulations
the advective electric field ($-\uu \times \textbf{B}$), associated with the bulk plasma motion in the
fan-plane and along the spine axis is
much stronger than its diffusive
counterpart (Figure \ref{fig:electric_field_a}), but since it is perpendicular to the magnetic field this
component causes no magnetic dissipation, and cannot be associated
with particle acceleration.

When determining
the values of the diffusive electric field one finds that the peak values in the
vicinity of the null-point are increasing with growing
displacement (see Figure \ref{fig:electric_field_b}), going from
initially zero to on the order of 50{\Vpm}, while in the fan-plane the
diffusive electric field reaches more than 90{\Vpm} at D = 2.95\,Mm.
However, as shown below (\textit{cf.}\ \Eq{cs-efield} and \Eq{cs-efield2}), to estimate the analogous
solar electric field, the simulation value should be reduced with a factor
equal to the power of 0.3 of the factor (about 20 for run 3) by which the boundary driving is
exaggerated; here we obtain $E \approx 90 / 20^{0.3}$, or about 36{\Vpm}.
Considering our more benign conditions, this is consistent with
\citet{1998SoPh..178..125P}, who find typical electric fields to be of
the order of 100 -- 300{\Vpm} under flaring conditions.
Electrons accelerated along the entire current sheet,
with an extent of about 15\,Mm (see Figure \ref{fig:j_fanplane}),
could nevertheless gain energies of up to about 300\,MeV in our case.
Of interest is also the shape of the average diffusive electric
field increase in the vicinity of the null-point, plotted in Figure
\ref{fig:electric_field_b} as a (*)-line. Its progression shows an almost
identical behavior as a plot of the increasing distance between
the inner and outer spine (see Figure \ref{fig:bfield}) plotted against
the displacement (plot not presented here).

\subsection{Comparison of stratified and non-stratified simulations}
\begin{figure*}[ht]
    \centering
    \includegraphics[scale=0.6]{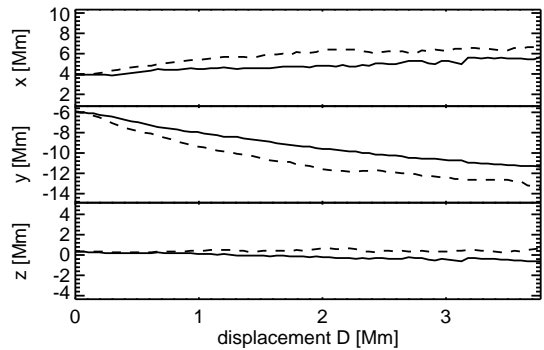}
    \caption{Null-point position in Mm in the coordinate system of the experiment. The vertical
    coordinate is $x$, and the two horizontal coordinates are $y$ and $z$.
The solid line shows the constant density run 4, while the dashed
line represents the stratified run 4S.}
    \label{fig:nullpointpos}
\end{figure*}

Since the null-point is an essential node of the magnetic skeleton, we
use it as a reference point for our investigation of the
influence of the density profile on the temporal changes of the
magnetic field. Here we compare simulation run 4S, which has a stratified
atmosphere, with simulation run 4, which has a constant
chromosphere-like density and temperature atmosphere (see, Table
\ref{tab:simulations}). Figure \ref{fig:nullpointpos} shows
the position of the null-point, connecting the fan-plane and
spine magnetic field lines,  in all three directions versus
the boundary displacement D.
In the stratified case the very dense plasma ($9\e{15}$ cm$^{-3}$)
at the bottom of the box gives rise to a low Alfv\'en speed (v$_{A}
= {B}/{\sqrt{\mu_0 \rho_m}}$), meaning that the higher density
impedes the propagation of the boundary disturbance into the box. But once
the disturbances reach beyond the transition region the ambient density
has fallen drastically, and the dense plasma from the lower atmosphere
starts spreading out faster
and pushes the null-point up. The velocity gradient induced by the
large density change causes strong dissipation and in some regions a disruption
into multiple smaller current sheets, which can also be
seen in the electric current density comparison in Figure
\ref{fig:j_condense_strati_compare.ps}, showing $|j|$ for runs 4 and
4S.

In the non-stratified runs the amount of plasma that moves upwards along the
magnetic field lines into the corona and contributes to pushing the null-point
upwards is much reduced, as seen in Figure \ref{fig:nullpointpos}.

For all runs we find a significant null-point motion (about
1--2\,Mm in $x$, 6\,Mm in $y$ and 1\,Mm in $z$) due to the applied driving
motion on the bottom boundary, which is comparable to the relative
motion of the inner spine.

\subsection{The influence of boundary conditions}
The driver of the magnetic field evolution is the boundary motion.
The magnetic field displacements imposed by the boundary motions
have a large influence on the spatial structure of the
magnetic skeleton, consisting of null-points, separatrix surfaces
(such as the fan-plane), separators, sources and flux domains
\citep[\textit{e.g.}][and references therein]{2008ApJ...675.1656P}. The
skeleton itself is a very robust structure,
which does not change from a topological point of view, but the detailed
appearance of it changes with boundary displacement, as already shown by
the analysis of the
null-point motion.

\begin{figure}
  \centering
    \subfloat[]{\label{fig:density_profile_a}\includegraphics[scale=0.4]{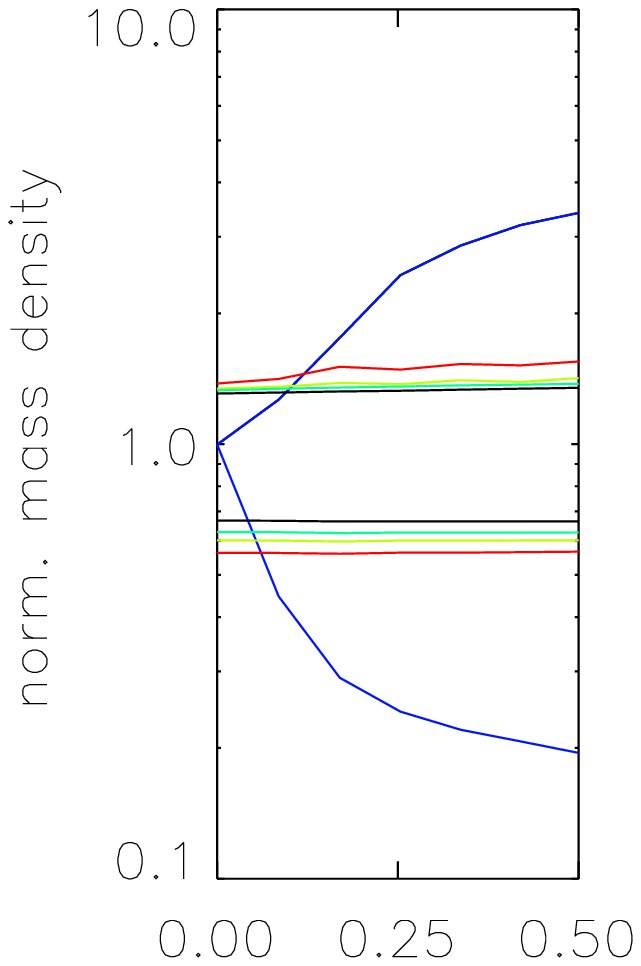}}\hspace{-1 cm}
    \subfloat[]{\label{fig:density_profile_b}\includegraphics[scale=0.4]{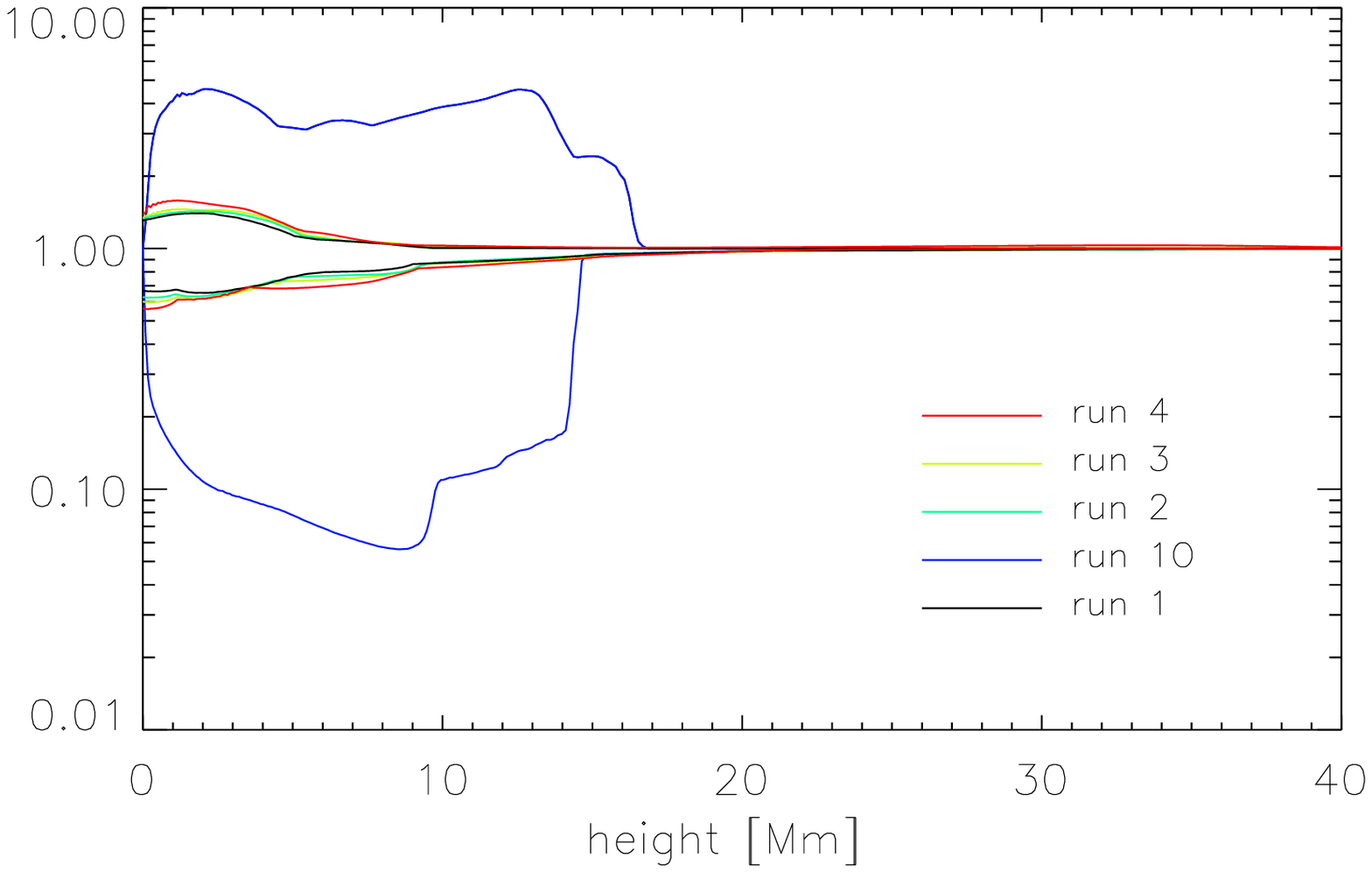}}
    \caption{Density profile min and max over each slice in height for the closed non-stratified
simulations, normalized to the asymptote: run 1, run 2, run 3 and
run 4 and the open non-stratified run 1O. (a) is
a zoom in of the first 0.5\,Mm of plot (b). The displacement of each run
is taken to be from the snapshot closest to D = 1.05\,Mm.}
    \label{fig:density_profile.ps}
\end{figure}

In Figure \ref{fig:density_profile_b} we compare the density profiles of the
closed runs 1, 2, 3 and 4 and the open boundary run 1O. The figure shows that all non-stratified
closed flow boundary runs develop a similar density profile: A certain expansion or
compression of a region connected to the closed boundary leads to the
same density profile, regardless of the driving speed.

The initial sound speed is about 83\,km\,s$^{-1}$ for the non-stratified runs,
while being much lower---on the order of 10\,km\,s$^{-1}$---in the lower parts
of the stratified runs. The sound speed influences among other things
the in- and outflows at open boundaries.

Figure \ref{fig:density_profile_a} illustrates that the closed boundary
conditions influence particularly the lowest density layer, where regions with
low and high density build up and cannot be emptied nor filled by plasma out-
and inflow. This is not very important for the dynamics of the system, but is
also not very solar-like. In the open case, there is a continuous mass exchange and
pressure equalization at the boundaries, in which case it is crucial that the sound speed is well
above the driving speed, so that the system has time to approach pressure balance.

The low sound speed in the stratified case poses a very tight restriction on
the driving speed, in order to avoid exaggerating the effects of inertia. On the
other hand there is a clear advantage of having stratification: it provides a
pool of mass for the corona to communicate with; the large amount of mass at
low temperature acts as a buffer, due to the low Alfv\'en and sound speed.

Overall, Figure \ref{fig:density_profile_b} confirms that the density contrast
is mainly caused by volume changes, which arise from the imposed boundary motions. If these
volume changes happen sufficiently slowly relative to the Alfv\'en and sound
speed, the driving speed loses its importance for the results (but not for
the computational cost of obtaining them!).

\subsection{Energy dissipation}
As the boundary moves according to the prescribed driving pattern, with
magnetic field lines passing through the boundary essentially `frozen in',
because of the boundary conditions, the system response may be mainly split
into two distinct components. The first is the change in potential magnetic
field energy due to the change of the vertical magnetic
field component brought about by the boundary motions; this is the smallest
amount by which the magnetic field energy could change. Secondly, in addition, the `free
magnetic energy' component will change as well. This non-potential part of the
magnetic field is (by
definition) associated with a non-zero electric current proportional
to $\nabla\times\mathbf{B}$. The electric current may either be smooth
and space-filling, or may be concentrated in electric current sheets,
corresponding to near-discontinuities of the magnetic field.

Formally, the rate of change of magnetic energy density $e_B = B^2 /2$ is
described by Equation (\ref{equ:em}), which shows that changes of magnetic energy
are due to the net effect of a (negative) divergence of the Poynting flux $F_P$,
conversion into bulk kinetic energy by the Lorentz work $W_L$, and
conversion to heat through Joule dissipation $Q_J$.
\begin{equation}
\frac{\partial e_{B}}{\partial t} = -\nabla\cdot \mathbf{F}_P - W_L - Q_J .\label{equ:em}
\end{equation}
The Poynting flux is defined as $\mathbf{F}_P = \mathbf{E} \times \mathbf{B}$,
and the Lorentz work is $W_L = \uu\cdot (\textbf{j} \times
\textbf{B})$. Joule dissipation, $Q_J = \mathbf{E}_{\eta}\cdot\mathbf{j} =
\eta\mathbf{j}^2$, primarily takes place in the strong current sheets.
Electric currents flow mainly
along the magnetic field in the corona and therefore $Q_J$ is a suitable
indicator for locations at which a significant component of the electric field
parallel to the magnetic field may exist.  Such a parallel electric field can
accelerate charged particles along the magnetic field
lines, resulting
\textit{e.g.} in the brightening of flare ribbons.

\begin{figure*}
    \centering
        \includegraphics[scale=0.46]{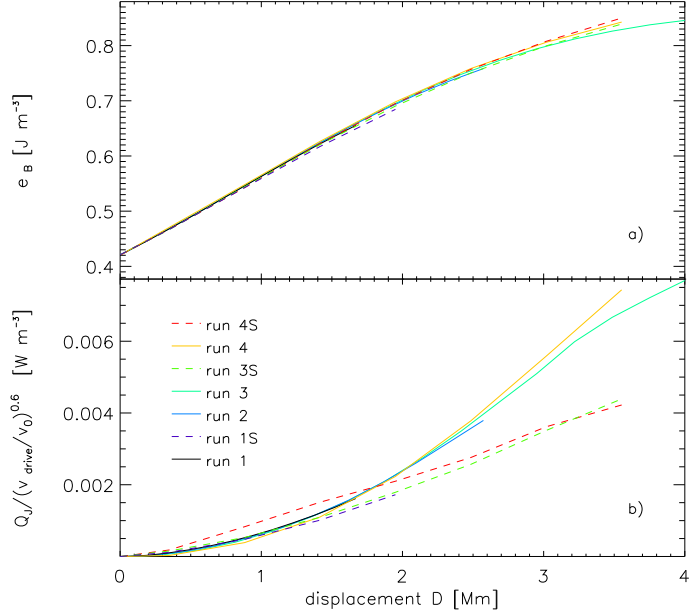}
    \caption{Comparison of the average magnetic energy a) and average magnetic
    dissipation divided by the normalized (v$_0$ = 10\,km s$^{-1}$) average boundary driving speed of each run b) for constant density runs (solid lines) and
    stratified runs (dashed lines) in the cutout of approximate size
    10 x 16 x 22\,Mm.}
    \label{fig:plot_joule_magnenergy.ps}
\end{figure*}

The evolutions of the magnetic energy and the Joule dissipation
normalized to the average driving speed are summarized in a plot covering
several simulation runs in Figure \ref{fig:plot_joule_magnenergy.ps}.
The first thing to notice is that the evolution of the magnetic
energy, when expressed in terms of the boundary displacement $D$,
is practically identical in all of the runs.
The reason for this is that most of the boundary work goes directly
into increasing the potential magnetic energy, while only a small amount goes into free magnetic energy.
From Figure \ref{fig:plot_joule_magnenergy.ps}b is seen that the dissipation
increases with increasing displacement. This indicates that
an increasing amount of free energy becomes available through the build
up of current structures in the null-point fan-plane as the experiment progresses.

A comparison between different power-law relations indicates, that the normalization of the magnetic dissipation by the average normalized
driving speed to the power 0.6 employed in Figure \ref{fig:plot_joule_magnenergy.ps}b
brings the curves showing the evolution
of magnetic dissipation for all the different runs closest together into a
relatively tight set of parallel relations. This illustrates that
the rate of magnetic dissipation, at any
given value of the displacement, is approximately proportional
to the rate of boundary displacement to the power 0.6.

The dissipation curves corresponding to the different experiments follow the same
general trend, although with some differences, in particular between
the stratified and non-stratified cases.
The dissipation is generally higher for the stratified cases than
for the non-stratified runs during early times and lower during late times.
The exception is run 1S, the run with the lowest driving speed, which
agrees closely with the non-stratified cases with the slowest driving speed.

The deviations from this common
asymptotic behavior are likely consequences of the low Alfv\'en speeds in the
dense layers of the stratified models, causing the dissipation in the stratified
runs to be initially high due to their higher densities at low
heights compared to the constant density cases. Later, when the motions
introduced by the driver reach greater heights, where the mass density is
lower than in the constant density runs, the stratified runs generally
display a lower dissipation.

We note in this context also that the viscous dissipation in the
system is much smaller than the Joule dissipation, as is
expected in a coronal environment.

The results summarized in Figure \ref{fig:plot_joule_magnenergy.ps}
illustrate that for a quantitatively accurate estimate of
properties related to the magnetic dissipation it is essential to drive in a
way which is compatible with the ordering of characteristic speeds
in the Sun; \textit{i.e.}, to keep the boundary speed smaller
than the Alfv\'en speed, and to scale the quantities down in proportion
to the speed-up factor used in the driving.

For the non-stratified runs we compute initial Alfv\'en speeds of
70 -- 1400\,km\,s$^{-1}$ at the
lower boundary, which is clearly higher than the driving speed of all
runs. So we expect, as is also shown by Figure \ref{fig:plot_joule_magnenergy.ps},
a similar dissipation increase with increasing displacement,
after some initial differences due to the driving speed differences.

\begin{figure}
    \centering
    \includegraphics[scale=0.14]{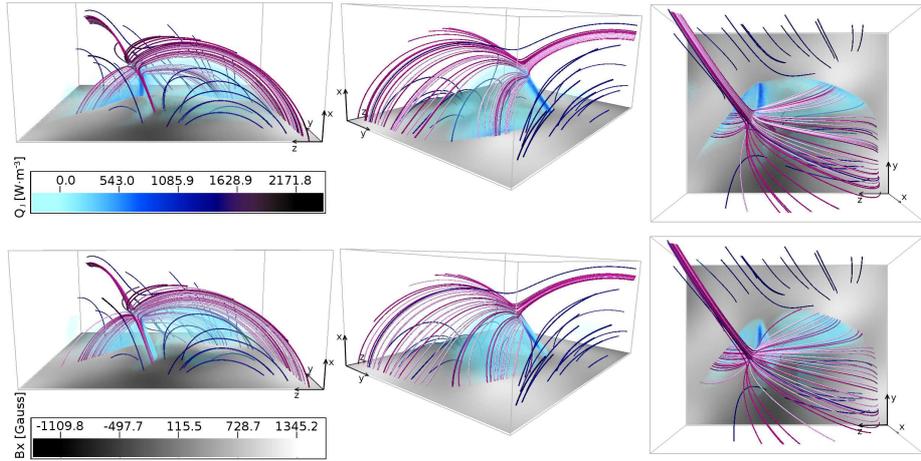}
    \caption{The highest joule dissipation of run 1 in the top row and
run 1S in the lower row is shown as a semitransparent volume rendering
at about D = 0.88\,Mm. The box size is $10\times16\times22$\,Mm. The
field lines represent magnetic field lines passing closest to the null
point (purple), in the close surrounding of the null-point (magenta)
and the overlying strong magnetic field (dark blue). The slice on the
lower boundary  shows the Bx magnetic field component. }
    \label{fig:dissipation_run4}
\end{figure}

In the stratified atmosphere runs the Alfv\'en speed increases with
height. At the lower boundary the initial Alfv\'en speed is approximately
5 -- 200\,km\,s$^{-1}$, where the minimum value falls below the driving speed
used in runs 3S and 4S. Run 1S is just at the edge of being
driven slower than the minimum Alfv\'en speed and indeed gives results
which are similar to the non-stratified case run 1. Figure
\ref{fig:dissipation_run4} shows that the stratified and non-stratified
cases are nevertheless distinguishable. The volume renderings show
the Joule dissipation normalized by the driving speed, for
run 1 (upper panel) and run 1S (lower panel). The locations of the dissipation
maxima agree nicely, but the dissipation maxima differ by a factor
of about 1.8, being higher in the stratified run 1S.

In summary, we find that the ratio of the driving speed to the
Alfv\'en speed has a noticeable impact on the dissipation level
and, as Figure
\ref{fig:plot_joule_magnenergy.ps} illustrates, that the stratified
simulations tend to display a progressive growth of deviations from the common
asymptotic relations defined by the non-stratified runs, unless the
driving speed is small compared to both the local Alfv\'en speed and
the sound speed.

However, these are relatively small deviations, compared to the
main trend, which is a proportionality between the magnetic
dissipation and the driving speed to the power 0.6.

\subsection{Scaling of magnetic dissipation in the current sheet}
As illustrated by Figure \ref{fig:fieldline_shear}, the magnetic field
line orientations on the two sides of the fan-plane only differ by
a small amount. This means that the electric current carried by
the current sheet, whose thickness (on the order of a few
grid cells in the present numerical model) we denote with $\Delta s$, is
much less than the maximal electric current density $2 B/\Delta s$ that
would result from a complete reversal of the magnetic field orientation
across the current sheet. The electric current density in the
current sheet (CS) is in fact on the order of
\begin{equation}
j_{{\rm CS}} \sim \Delta B \; \Delta s^{-1} \approx sin(\phi) B_{{\rm CS}}
\Delta s^{-1} \approx \phi B_{{\rm CS}} \Delta s^{-1} , \label{eq:cs-current}
\end{equation}
for small $\phi$,
where $B_{{\rm CS}}$ is a typical strength of the magnetic field just outside the
current sheet, and $\phi$ is an angle characterizing the difference
of direction of field lines on the two sides of the current sheet.

A fundamental question is now how the total dissipation in the current
sheet, and hence the average rate of reconnection in the structure,
depends on factors such as the numerical resolution and rate of
work done at the boundary. By construction the code keeps current
sheets just barely resolved.  A change in numerical resolution is thus
directly mapped into a proportional change of the current sheet thickness
$\Delta s$. This behavior is obtained by making $\eta$ essentially
proportional to the grid size (\textit{cf.}\ \Eq{eta}).
To a first order approximation both $B_{{\rm CS}}$ and
$\phi$ are independent of $\Delta s$.
By \Eq{cs-current} the
electric current density is inversely proportional to $\Delta s$, and
the magnetic dissipation rate per unit volume $Q_J$ therefore scales
as
\begin{equation}\label{eq:cs-dissipation}
Q_{{\rm CS}} = \eta j_{CS}^2 \sim \Delta B^2 \Delta s^{-1} = \phi^2 B_{CS}^2  \Delta s^{-1}.
\label{eq:dissipation_CS}
\end{equation}
To obtain the total dissipation in the current sheet this needs to
be multiplied with the volume of the current sheet,
\begin{equation}\label{eq:cs-volume}
V_{{\rm CS}} \sim A_{{\rm CS}}\Delta s,
\end{equation}
where we denote by $A_{{\rm CS}}$ the total area of the current sheet.
We thus conclude that the total dissipation is
\begin{equation}\label{eq:cs-dissipation2}
Q_{{\rm CS}} V_{{\rm CS}} \sim \Delta B^2 A_{{\rm CS}} ,
\end{equation}
and hence is, to lowest order, {\em independent of} $\Delta s$ and the resistivity in the current sheet.
Note that, as a consequence of $\phi$ being small, reconnection
in the current sheet can proceed without
requiring super-Alfv\'enic outflow velocities from the current
sheet.

Estimating now the diffusive electric field in the current sheet
we find that it scales as
\begin{equation}\label{eq:cs-efield}
E_{{\rm CS}} = \eta_{{\rm CS}} j_{{\rm CS}} \propto \Delta B = \phi B_{{\rm CS}},
\label{eq:resistive_efield}
\end{equation}
again independent of $\Delta s$, but proportional to the
change of magnetic field direction across the current sheet and hence
proportional to $\phi$.

Generally the work done by the boundary must
go into an increase in magnetic energy (potential plus free magnetic
energy), or into kinetic energy or ohmic dissipation.
In the present case the magnetic dissipation
is able to nearly keep up with the free energy input, and the system
essentially goes through a series of states not far from potential.
As Figure \ref{fig:plot_joule_magnenergy.ps} shows, the total
magnetic energy depends mainly on the displacement and very little on
the driving speed itself, while the dissipation is essentially
proportional to how fast we drive at the boundary to the power of 0.6, thus $Q_{{\rm CS}}
\propto v^{0.6}$.
So we conclude with the help of Equations (\ref{eq:dissipation_CS}) and (\ref{eq:resistive_efield}) that
the magnitude of the parallel
electric field along the current sheet scales as
\begin{equation}\label{eq:cs-efield2}
E_{{\rm CS}} \propto \phi \propto v^{0.3} .
\end{equation}
This electric field  -- driving speed relation is not an artifact of the
chosen numerical method; the scaling of $\eta$ with $\Delta s$ is
generic to all numerical methods, and serves to ensure that higher
numerical resolution can be used to reach larger magnetic Reynolds
and Lundquist numbers.

With much higher rates of stressing, or with much higher numerical
resolution, the current sheet may need to fragment and enter a
turbulent regime, in order to support the required amounts of
dissipation and reconnection in the face of increasing constraints
by the thinness of the current sheets. This is a process that is by
now well understood to be able to take over when the need arises
\citep{1996JGR...10113445G,1997LNP...489..179N,%
2005ApJ...618.1020G,2009ApJ...705..347B,2011A&A...530A.112B,%
2012ApJ...747..109N,2011AdSpR..47.1508P}.

We thus conclude that, provided that conditions that apply equally
to both numerical simulations and the Sun are fulfilled, neither
the total magnetic dissipation in a current sheet structure, nor
the diffusive part of the electric field depend, to lowest order,
on the electrical resistivity---or for that matter on the precise
mechanism that sets the level of the electrical resistivity.

\subsection{Triggering of rapid energy release}
As in \citet{2009ApJ...700..559M} there is no
significant and sudden relaxation of the system, with an
energy release that could correspond to a flare, even though the
simulations cover enough solar time to get across the observed
flaring event. Looking at the potential part of the magnetic field
at different displacement steps in the
simulations, we find steadily increasing magnetic potential energies,
and most of the build-up of magnetic energy seen in the experiment
actually goes into increasing the
potential rather than the non-potential part of the magnetic energy.
As discussed above, the magnetic dissipation is able to keep up with
varying levels of boundary work, with a residual amount of free energy
scaling, if it is proportional to the rate of dissipation,
approximately as the driving speed raised to 0.6. The stress in
the system is demonstrably moderate, even with our exaggerated driving
speed, and we therefore cannot expect boundary motions of the type
applied here to be able to explain
violent events similar to the observed solar flare.

We consider four possibilities for this difference in behavior
between these MHD simulations and the Sun:
\begin{enumerate}
\item The limited resolution of the numerical experiment is preventing an
instability from occurring that would otherwise trigger a flare-like event.
\item A flare-like event would occur if taking into account
kinetic effects (\textit{e.g.} by using particle-in-cell simulations).
\item A flare-like event could take place, with MHD alone, but would
require an additional Poynting flux through the boundary, in addition
to the Poynting flux generated by the simple driver implemented here
and in \citet{2009ApJ...700..559M}.
\item The additional free energy available because the system
was initially not in a potential state could have helped.
\end{enumerate}

As demonstrated above, numerical resolution does not to a first order determine the level
of dissipation in current sheets, once they are reasonably well resolved.
Indeed, by the arguments in the preceding subsection we expect the
level of stress (as measured for example by the angle $\phi$)
caused by the type of motions we employ in the current investigation
to be {\em smaller} in the Sun (by a factor of \textit{e.g.}\ $\approx
7^{0.3} \approx 2$ relative to our runs 1, 1O and 1S), and it is
thus very unlikely that the flare was triggered by accumulation of
stress from this particular type of boundary motion.

An MHD-instability could in principle occur at a later point in time than
to where our runs go; but as demonstrated by our experiments even a
driving speed that is highly exaggerated relative to the solar value is
not able to build up sufficient free energy to account for a C-class
flare: The maximum \textit{total} magnetic energy in the entire simulated domain of our experiment is
on the order of $2\e{30}$\,ergs, of which only a very small fraction is free
energy. Estimates of C-flare emission are larger than even our potential energy
\citep{2011A&A...530A..84K}, and hence the free energy available in our
model is not sufficient to power a C-class flare.
As discussed above, the level of stress is expected to be
proportional to the driving speed to some small positive power, and is expected to be
largely independent of the level of resistivity, and hence it is
unlikely that more stress would build up in the Sun than in these
numerical simulations.

\begin{figure}
    \centering
        \includegraphics[scale=0.7]{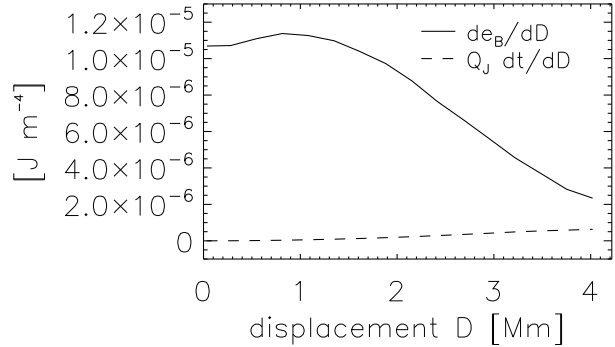}
    \caption{The average rate of change of magnetic energy for run 3 is
    plotted as a solid line, and the dissipation is
    shown as a dashed line. Both are restricted to the cutout
    of approximate size 10 x 16 x 22\,Mm used previously.}
       \label{fig:plot_joule_magnenergy_time.ps}
\end{figure}

Figure \ref{fig:plot_joule_magnenergy_time.ps}, which is
representative of all runs, illustrates that the rate of change of magnetic
energy increases in the beginning, as a result of the work done by the
lower boundary. This energy input loses efficiency as the angle between
the field lines and the driving boundary approaches 90 degrees.
Additionally, since the driving pattern location is fixed, the applied
stress becomes less and less efficient, as the flux in this particular
region is gradually removed. The dissipation is small compared to the rate
of change of magnetic energy, implying that the dissipated
free energy only makes up a tiny fraction compared to
the change of the potential energy. This is in agreement with Figure
\ref{fig:fieldline_shear},
which illustrates that the shear of the magnetic field lines in the
current sheet is not very large, even with our exaggerated driving speed.
Hence the reconnection is rapid enough to keep the boundary work at
a given displacement at an approximately constant level, and to keep
the system in a near potential state. We are thus far away from an instability
to occur and it is improbable to achieve one at a later point of the experiment
with the applied boundary motion. The mostly compressive boundary
driving pattern probably does not represent reality accurately enough, and
additional shearing, twisting, or emerging motions may have been present.

By similar arguments kinetic effects are not expected to
have a major influence on the macroscopic dissipation behavior and
reconnection rate. This is consistent with the findings by
\citet{2005GeoRL..3206105B}, which show that the same amounts of energy release take
place for the same type of reconnection event, independent of the dissipation
mechanism. It also shows that there are only small differences in
the time scales of the event going from MHD to PIC simulations.
These results are entirely consistent with the current sheet scaling
arguments presented above, since they hold true independent of the
nature of the dissipation mechanism.

We thus are left with the option that the additional free energy
mentioned in the third and forth alternative above is the most likely
explanation for the observed solar flare. This conclusion is consistent with
the fact that all observational evidence associates flaring either with new
flux emergence, or with magnetic configurations that evolve into states that
are decidedly MHD-unstable; \textit{i.e.}, ones in which lower energy states become
accessible and where dissipative effects make it possible to reach those lower
energy states and dynamic instabilities therefore develop. Some amount of
free magnetic energy was clearly present in the system at the
time where we are forced by circumstances to assume a potential initial state.
Whether that extra free energy alone would have been enough to cause a flare
several hours later is an open question. Our driver pattern might have been
able to add slightly more magnetic energy to a configuration that already
contained free magnetic energy, but that effect is probably small.  We
consider the addition of Poynting flux, increasing the free magnetic
energy of the system at a considerably higher rate than can be achieved
with our driver pattern as the more likely alternative. Additional Poynting
flux across the lower boundary can take the form of either emerging flux
(Poynting flux due to vertical velocity) or twisting boundary motions
(Poynting flux due to vortical horizontal velocity).

As evidence for this view,
we consider magnetograms starting on November 12, 2002 in the early afternoon,
when a strong negative polarity first appeared within the positive polarity.
The negative polarity expanded and several eruptions could be seen at EUV
wavelengths. During the interval simulated here one observes an additional
increase in the negative flux inside the dome, and there are also other
significant rearrangements that our driver pattern is not able to represent or explain.

\section{Conclusions}
\label{sec:conclusions}
The main topic of the presented work is a study of the influence of the
driving speed for stratified and non-stratified atmosphere models used to
simulate 3D reconnection events in the solar corona. The major findings are:

\begin{itemize}
\item Sufficiently low driving speeds lead to similar plasma
behavior, including similar evolution of the magnetic energy density
(after compensating for the expected effects
of the different driving speeds).

\item The magnetic dissipation and the diffusive electric field ---
capable of accelerating charged particles depend only weakly on
the grid resolution in the numerical experiments.

\item When driving is taking place sufficiently slowly, the rate of magnetic
dissipation increases approximately as the boundary driving speed raised to
the power 0.6 while the diffusive electric field increases approximately
as the driving speed raised to the power 0.3.

\item The driving speed has a larger impact on the general plasma behavior
in the cases with stratified atmospheres, while its influence is minor for
non-stratified runs with closed flow boundary conditions, if compared at
the same boundary motion displacements.

\item The sound speed compared to the driving speed determines the exchange of
plasma at the boundaries. Voids can be filled in open boundary runs, while
pressure differences at the boundary are carried along in the closed
boundary cases. This is not an issue in the stratified runs, due to the
large reservoirs of relatively cold gas in the vicinity of the lower
boundary. On the other hand, the lower information speeds there set
severe restrictions on the driving speed.
\end{itemize}

Our investigations suggest that the applied simple driver pattern is unlikely to be
able to cause a flare-like energy release in the simulations (as well as in
the Sun). In fact, based on a comparison of the available free energy
in our model with estimates from observations of C-class flares we anticipate
that the corresponding solar configuration must have had significantly more
free magnetic energy added through the boundary, or must otherwise have been
in a strongly non-potential state already before the studied time interval.

\begin{acks}
We would like to especially thank Jacob Trier Frederiksen and Troels
Haugb{\o}lle for valuable discussions and for their assistance with the
simulations.  We thank Guillaume Aulanier and Sophie Masson
for providing us with their MHD data and driver information.
We also thank the referee for useful comments and criticism.
This work has been supported by the Niels Bohr International Academy and the
SOLAIRE Research Training Network of the European Commission
(MRTN-CT-2006-035484). The work of \AA N was partially supported
by the Danish Research Council for Independent Research (FNU) and the funding from the European Commission's Seventh Framework Programme (FP7/2007-2013) under the grant agreement SWIFF (project n° 263340, www.swiff.eu).
The data for the magnetic field extrapolation were taken
from the SOHO catalog. SOHO is a project of international cooperation
between ESA and NASA. We furthermore acknowledge that the results in this
paper have been achieved using resources at the Danish Center for
Scientific Computing in Copenhagen, as well as PRACE and GCS/NIC Research Infrastructure
resources on JUGENE and JUROPA based at J\"ulich in Germany.

\end{acks}

  \bibliographystyle{spr-mp-sola}
  \bibliography{adhoc,references}

\begin{thebibliography}{27}
\ifx \bisbn   \undefined \def \bisbn  #1{ISBN #1}\fi
\ifx \binits  \undefined \def \binits#1{#1}\fi
\ifx \bauthor  \undefined \def \bauthor#1{#1}\fi
\ifx \batitle  \undefined \def \batitle#1{#1}\fi
\ifx \bjtitle  \undefined \def \bjtitle#1{\textit{#1}}\fi
\ifx \bvolume  \undefined \def \bvolume#1{\textbf{#1}}\fi
\ifx \byear  \undefined \def \byear#1{#1}\fi
\ifx \bissue  \undefined \def \bissue#1{#1}\fi
\ifx \bfpage  \undefined \def \bfpage#1{#1}\fi
\ifx \blpage  \undefined \def \blpage #1{#1}\fi
\ifx \burl  \undefined \def \burl#1{\textsf{#1}}\fi
\ifx \href  \undefined \def \href#1#2{\textsf{#2}}\fi
\ifx \doiurl  \undefined \def
  \doiurl#1{\href{http://dx.doi.org/#1}{\textsf{#1}}}\fi
\ifx \betal  \undefined \def \betal{\textit{et al.}}\fi
\ifx \binstitute  \undefined \def \binstitute#1{#1}\fi
\ifx \bctitle  \undefined \def \bctitle#1{#1}\fi
\ifx \beditor  \undefined \def \beditor#1{#1}\fi
\ifx \bpublisher  \undefined \def \bpublisher#1{#1}\fi
\ifx \bbtitle  \undefined \def \bbtitle#1{\textit{#1}}\fi
\ifx \bedition  \undefined \def \bedition#1{#1}\fi
\ifx \bseriesno  \undefined \def \bseriesno#1{\textbf{#1}}\fi
\ifx \blocation  \undefined \def \blocation#1{#1}\fi
\ifx \bsertitle  \undefined \def \bsertitle#1{\textit{#1}}\fi
\ifx \bsnm \undefined \def \bsnm#1{#1}\fi
\ifx \bsuffix \undefined \def \bsuffix#1{#1}\fi
\ifx \bparticle \undefined \def \bparticle#1{#1}\fi
\ifx \barticle \undefined \def \barticle#1{}\fi
\ifx \botherref \undefined \def \botherref#1{}\fi
\ifx \url \undefined \def \url#1{\textsf{#1}}\fi
\ifx \bchapter \undefined \def \bchapter#1{}\fi
\ifx \bbook \undefined \def \bbook#1{}\fi
\ifx \bcomment \undefined \def \bcomment#1{#1}\fi
\ifx \oauthor \undefined \def \oauthor#1{#1}\fi
\ifx \citeauthoryear \undefined \def \citeauthoryear#1{#1}\fi
\def \endbibitem {}
\ifx \bconflocation  \undefined \def \bconflocation#1{#1} \fi

\bibitem[\protect\citeauthoryear{{Archontis}
  \textit{et~al.}}{2004}]{2004A&A...426.1047A}
\begin{barticle}
\bauthor{\bsnm{{Archontis}}, \binits{V.}},
\bauthor{\bsnm{{Moreno-Insertis}}, \binits{F.}},
\bauthor{\bsnm{{Galsgaard}}, \binits{K.}},
\bauthor{\bsnm{{Hood}}, \binits{A.}},
\bauthor{\bsnm{{O'Shea}}, \binits{E.}}:
\byear{2004},
\batitle{{Emergence of magnetic flux from the convection zone into the
  corona}}.
\bjtitle{\aap}
\bvolume{426},
\bfpage{1047}\,--\,\blpage{1063}.
doi:\doiurl{10.1051/0004-6361:20035934}.
\end{barticle}
\endbibitem

\bibitem[\protect\citeauthoryear{{Arzner} and
  {Vlahos}}{2006}]{2006A&A...454..957A}
\begin{barticle}
\bauthor{\bsnm{{Arzner}}, \binits{K.}},
\bauthor{\bsnm{{Vlahos}}, \binits{L.}}:
\byear{2006},
\batitle{{Gyrokinetic electron acceleration in the force-free corona with
  anomalous resistivity}}.
\bjtitle{\aap}
\bvolume{454},
\bfpage{957}\,--\,\blpage{967}.
doi:\doiurl{10.1051/0004-6361:20064953}.
\end{barticle}
\endbibitem

\bibitem[\protect\citeauthoryear{{Berger} and
  {Asgari-Targhi}}{2009}]{2009ApJ...705..347B}
\begin{barticle}
\bauthor{\bsnm{{Berger}}, \binits{M.A.}},
\bauthor{\bsnm{{Asgari-Targhi}}, \binits{M.}}:
\byear{2009},
\batitle{{Self-organized Braiding and the Structure of Coronal Loops}}.
\bjtitle{\apj}
\bvolume{705},
\bfpage{347}\,--\,\blpage{355}.
doi:\doiurl{10.1088/0004-637X/705/1/347}.
\end{barticle}
\endbibitem

\bibitem[\protect\citeauthoryear{{Bingert} and
  {Peter}}{2011}]{2011A&A...530A.112B}
\begin{barticle}
\bauthor{\bsnm{{Bingert}}, \binits{S.}},
\bauthor{\bsnm{{Peter}}, \binits{H.}}:
\byear{2011},
\batitle{{Intermittent heating in the solar corona employing a 3D MHD model}}.
\bjtitle{\aap}
\bvolume{530},
\bfpage{A112}.
doi:\doiurl{10.1051/0004-6361/201016019}.
\end{barticle}
\endbibitem

\bibitem[\protect\citeauthoryear{{Birn}
  \textit{et~al.}}{2005}]{2005GeoRL..3206105B}
\begin{barticle}
\bauthor{\bsnm{{Birn}}, \binits{J.}},
\bauthor{\bsnm{{Galsgaard}}, \binits{K.}},
\bauthor{\bsnm{{Hesse}}, \binits{M.}},
\bauthor{\bsnm{{Hoshino}}, \binits{M.}},
\bauthor{\bsnm{{Huba}}, \binits{J.}},
\bauthor{\bsnm{{Lapenta}}, \binits{G.}},
\bauthor{\bsnm{{Pritchett}}, \binits{P.L.}},
\bauthor{\bsnm{{Schindler}}, \binits{K.}},
\bauthor{\bsnm{{Yin}}, \binits{L.}},
\bauthor{\bsnm{{B{\"u}chner}}, \binits{J.}},
\bauthor{\bsnm{{Neukirch}}, \binits{T.}},
\bauthor{\bsnm{{Priest}}, \binits{E.R.}}:
\byear{2005},
\batitle{{Forced magnetic reconnection}}.
\bjtitle{\grl}
\bvolume{32},
\bfpage{6105}.
doi:\doiurl{10.1029/2004GL022058}.
\end{barticle}
\endbibitem

\bibitem[\protect\citeauthoryear{{Fan} and
  {Gibson}}{2003}]{2003ApJ...589L.105F}
\begin{barticle}
\bauthor{\bsnm{{Fan}}, \binits{Y.}},
\bauthor{\bsnm{{Gibson}}, \binits{S.E.}}:
\byear{2003},
\batitle{{The Emergence of a Twisted Magnetic Flux Tube into a Preexisting
  Coronal Arcade}}.
\bjtitle{\apjl}
\bvolume{589},
\bfpage{L105}\,--\,\blpage{L108}.
doi:\doiurl{10.1086/375834}.
\end{barticle}
\endbibitem

\bibitem[\protect\citeauthoryear{{Galsgaard} and
  {Nordlund}}{1996}]{1996JGR...10113445G}
\begin{barticle}
\bauthor{\bsnm{{Galsgaard}}, \binits{K.}},
\bauthor{\bsnm{{Nordlund}}, \binits{{\AA}.}}:
\byear{1996},
\batitle{{Heating and activity of the solar corona 1. Boundary shearing of an
  initially homogeneous magnetic field}}.
\bjtitle{\jgr}
\bvolume{101},
\bfpage{13445}\,--\,\blpage{13460}.
doi:\doiurl{10.1029/96JA00428}.
\end{barticle}
\endbibitem

\bibitem[\protect\citeauthoryear{{Galsgaard} and
  {Pontin}}{2011}]{2011A&A...529A..20G}
\begin{barticle}
\bauthor{\bsnm{{Galsgaard}}, \binits{K.}},
\bauthor{\bsnm{{Pontin}}, \binits{D.I.}}:
\byear{2011},
\batitle{{Steady state reconnection at a single 3D magnetic null point}}.
\bjtitle{\aap}
\bvolume{529},
\bfpage{A20}.
doi:\doiurl{10.1051/0004-6361/201014359}.
\end{barticle}
\endbibitem

\bibitem[\protect\citeauthoryear{Green}{1989}]{Green89}
\begin{barticle}
\bauthor{\bsnm{Green}, \binits{J.M.}}:
\byear{1989},
\batitle{Geometrical properties of 3d reconnecting magnetic fields with nulls}.
\bjtitle{\jgr}
\bvolume{93},
\bfpage{8583}.
\end{barticle}
\endbibitem

\bibitem[\protect\citeauthoryear{{Gudiksen} and
  {Nordlund}}{2002}]{2002ApJ...572L.113G}
\begin{barticle}
\bauthor{\bsnm{{Gudiksen}}, \binits{B.V.}},
\bauthor{\bsnm{{Nordlund}}, \binits{{\AA}.}}:
\byear{2002},
\batitle{{Bulk Heating and Slender Magnetic Loops in the Solar Corona}}.
\bjtitle{\apjl}
\bvolume{572},
\bfpage{L113}\,--\,\blpage{L116}.
doi:\doiurl{10.1086/341600}.
\end{barticle}
\endbibitem

\bibitem[\protect\citeauthoryear{{Gudiksen} and
  {Nordlund}}{2005}]{2005ApJ...618.1020G}
\begin{barticle}
\bauthor{\bsnm{{Gudiksen}}, \binits{B.V.}},
\bauthor{\bsnm{{Nordlund}}, \binits{{\AA}.}}:
\byear{2005},
\batitle{{An Ab Initio Approach to the Solar Coronal Heating Problem}}.
\bjtitle{\apj}
\bvolume{618},
\bfpage{1020}\,--\,\blpage{1030}.
doi:\doiurl{10.1086/426063}.
\end{barticle}
\endbibitem

\bibitem[\protect\citeauthoryear{{Hyman}}{1979}]{1979acmp.proc..313H}
\begin{bchapter}
\bauthor{\bsnm{{Hyman}}, \binits{J.M.}}:
\byear{1979},
\bctitle{{A method of lines approach to the numerical solution of conservation
  laws}}.
In: \bbtitle{Advances in Computer Methods for Partial Differential Equations -
  III},
\bfpage{313}\,--\,\blpage{321}.
\end{bchapter}
\endbibitem

\bibitem[\protect\citeauthoryear{{Kretzschmar}}{2011}]{2011A&A...530A..84K}
\begin{barticle}
\bauthor{\bsnm{{Kretzschmar}}, \binits{M.}}:
\byear{2011},
\batitle{{The Sun as a star: observations of white-light flares}}.
\bjtitle{\aap}
\bvolume{530},
\bfpage{A84}.
doi:\doiurl{10.1051/0004-6361/201015930}.
\end{barticle}
\endbibitem

\bibitem[\protect\citeauthoryear{{Kritsuk}
  \textit{et~al.}}{2011}]{2011ApJ...737...13K}
\begin{barticle}
\bauthor{\bsnm{{Kritsuk}}, \binits{A.G.}},
\bauthor{\bsnm{{Nordlund}}, \binits{{\AA}.}},
\bauthor{\bsnm{{Collins}}, \binits{D.}},
\bauthor{\bsnm{{Padoan}}, \binits{P.}},
\bauthor{\bsnm{{Norman}}, \binits{M.L.}},
\bauthor{\bsnm{{Abel}}, \binits{T.}},
\bauthor{\bsnm{{Banerjee}}, \binits{R.}},
\bauthor{\bsnm{{Federrath}}, \binits{C.}},
\bauthor{\bsnm{{Flock}}, \binits{M.}},
\bauthor{\bsnm{{Lee}}, \binits{D.}},
\bauthor{\bsnm{{Li}}, \binits{P.S.}},
\bauthor{\bsnm{{M{\"u}ller}}, \binits{W.-C.}},
\bauthor{\bsnm{{Teyssier}}, \binits{R.}},
\bauthor{\bsnm{{Ustyugov}}, \binits{S.D.}},
\bauthor{\bsnm{{Vogel}}, \binits{C.}},
\bauthor{\bsnm{{Xu}}, \binits{H.}}:
\byear{2011},
\batitle{{Comparing Numerical Methods for Isothermal Magnetized Supersonic
  Turbulence}}.
\bjtitle{\apj}
\bvolume{737},
\bfpage{13}.
doi:\doiurl{10.1088/0004-637X/737/1/13}.
\end{barticle}
\endbibitem

\bibitem[\protect\citeauthoryear{{Masson}
  \textit{et~al.}}{2009}]{2009ApJ...700..559M}
\begin{barticle}
\bauthor{\bsnm{{Masson}}, \binits{S.}},
\bauthor{\bsnm{{Pariat}}, \binits{E.}},
\bauthor{\bsnm{{Aulanier}}, \binits{G.}},
\bauthor{\bsnm{{Schrijver}}, \binits{C.J.}}:
\byear{2009},
\batitle{{The Nature of Flare Ribbons in Coronal Null-Point Topology}}.
\bjtitle{\apj}
\bvolume{700},
\bfpage{559}\,--\,\blpage{578}.
doi:\doiurl{10.1088/0004-637X/700/1/559}.
\end{barticle}
\endbibitem

\bibitem[\protect\citeauthoryear{{Ng}, {Lin}, and
  {Bhattacharjee}}{2012}]{2012ApJ...747..109N}
\begin{barticle}
\bauthor{\bsnm{{Ng}}, \binits{C.S.}},
\bauthor{\bsnm{{Lin}}, \binits{L.}},
\bauthor{\bsnm{{Bhattacharjee}}, \binits{A.}}:
\byear{2012},
\batitle{{High-Lundquist Number Scaling in Three-dimensional Simulations of
  Parker's Model of Coronal Heating}}.
\bjtitle{\apj}
\bvolume{747},
\bfpage{109}.
doi:\doiurl{10.1088/0004-637X/747/2/109}.
\end{barticle}
\endbibitem

\bibitem[\protect\citeauthoryear{{Nordlund} and
  {Galsgaard}}{1997a}]{Nordlund1997}
\begin{botherref}
\oauthor{\bsnm{{Nordlund}}, \binits{A.}},
\oauthor{\bsnm{{Galsgaard}}, \binits{K.}}:
1997a,
{A 3D MHD code for Parallel Computers}.
Technical report,
{Niels Bohr Institute}.
\end{botherref}
\endbibitem

\bibitem[\protect\citeauthoryear{{Nordlund} and
  {Galsgaard}}{1997b}]{1997LNP...489..179N}
\begin{bchapter}
\bauthor{\bsnm{{Nordlund}}, \binits{A.}},
\bauthor{\bsnm{{Galsgaard}}, \binits{K.}}:
\byear{1997}b,
\bctitle{{Topologically Forced Reconnection}}.
In: \beditor{\bsnm{{Simnett}}, \binits{G.M.}},
\beditor{\bsnm{{Alissandrakis}}, \binits{C.E.}},
\beditor{\bsnm{{Vlahos}}, \binits{L.}} (eds.)
\bbtitle{European Meeting on Solar Physics},
\bsertitle{Lecture Notes in Physics, Berlin Springer Verlag}
\bseriesno{489},
\bfpage{179}.
doi:\doiurl{10.1007/BFb0105676}.
\end{bchapter}
\endbibitem

\bibitem[\protect\citeauthoryear{{Parnell}, {Haynes}, and
  {Galsgaard}}{2008}]{2008ApJ...675.1656P}
\begin{barticle}
\bauthor{\bsnm{{Parnell}}, \binits{C.E.}},
\bauthor{\bsnm{{Haynes}}, \binits{A.L.}},
\bauthor{\bsnm{{Galsgaard}}, \binits{K.}}:
\byear{2008},
\batitle{{Recursive Reconnection and Magnetic Skeletons}}.
\bjtitle{\apj}
\bvolume{675},
\bfpage{1656}\,--\,\blpage{1665}.
doi:\doiurl{10.1086/527532}.
\end{barticle}
\endbibitem

\bibitem[\protect\citeauthoryear{{Parnell}
  \textit{et~al.}}{1996}]{1996PhPl....3..759P}
\begin{barticle}
\bauthor{\bsnm{{Parnell}}, \binits{C.E.}},
\bauthor{\bsnm{{Smith}}, \binits{J.M.}},
\bauthor{\bsnm{{Neukirch}}, \binits{T.}},
\bauthor{\bsnm{{Priest}}, \binits{E.R.}}:
\byear{1996},
\batitle{{The structure of three-dimensional magnetic neutral points}}.
\bjtitle{Physics of Plasmas}
\bvolume{3},
\bfpage{759}\,--\,\blpage{770}.
doi:\doiurl{10.1063/1.871810}.
\end{barticle}
\endbibitem

\bibitem[\protect\citeauthoryear{{Pontin}}{2011}]{2011AdSpR..47.1508P}
\begin{barticle}
\bauthor{\bsnm{{Pontin}}, \binits{D.I.}}:
\byear{2011},
\batitle{{Three-dimensional magnetic reconnection regimes: A review}}.
\bjtitle{Advances in Space Research}
\bvolume{47},
\bfpage{1508}\,--\,\blpage{1522}.
doi:\doiurl{10.1016/j.asr.2010.12.022}.
\end{barticle}
\endbibitem

\bibitem[\protect\citeauthoryear{{Pontin}, {Bhattacharjee}, and
  {Galsgaard}}{2007}]{2007PhPl...14e2106P}
\begin{barticle}
\bauthor{\bsnm{{Pontin}}, \binits{D.I.}},
\bauthor{\bsnm{{Bhattacharjee}}, \binits{A.}},
\bauthor{\bsnm{{Galsgaard}}, \binits{K.}}:
\byear{2007},
\batitle{{Current sheet formation and nonideal behavior at three-dimensional
  magnetic null points}}.
\bjtitle{Physics of Plasmas}
\bvolume{14}(\bissue{5}),
\bfpage{052106}.
doi:\doiurl{10.1063/1.2722300}.
\end{barticle}
\endbibitem

\bibitem[\protect\citeauthoryear{{Priest} and
  {Titov}}{1996}]{1996RSPTA.354.2951P}
\begin{barticle}
\bauthor{\bsnm{{Priest}}, \binits{E.R.}},
\bauthor{\bsnm{{Titov}}, \binits{V.S.}}:
\byear{1996},
\batitle{{Magnetic Reconnection at Three-Dimensional Null Points}}.
\bjtitle{Royal Society of London Philosophical Transactions Series A}
\bvolume{354},
\bfpage{2951}\,--\,\blpage{2992}.
doi:\doiurl{10.1098/rsta.1996.0136}.
\end{barticle}
\endbibitem

\bibitem[\protect\citeauthoryear{{Priest}, {Bungey}, and
  {Titov}}{1997}]{1997GApFD..84..127P}
\begin{barticle}
\bauthor{\bsnm{{Priest}}, \binits{E.R.}},
\bauthor{\bsnm{{Bungey}}, \binits{T.N.}},
\bauthor{\bsnm{{Titov}}, \binits{V.S.}}:
\byear{1997},
\batitle{{The 3D topology and interaction of complex magnetic flux systems}}.
\bjtitle{Geophysical and Astrophysical Fluid Dynamics}
\bvolume{84},
\bfpage{127}\,--\,\blpage{163}.
doi:\doiurl{10.1080/03091929708208976}.
\end{barticle}
\endbibitem

\bibitem[\protect\citeauthoryear{{Pudovkin}
  \textit{et~al.}}{1998}]{1998SoPh..178..125P}
\begin{barticle}
\bauthor{\bsnm{{Pudovkin}}, \binits{M.I.}},
\bauthor{\bsnm{{Zaitseva}}, \binits{S.A.}},
\bauthor{\bsnm{{Shumilov}}, \binits{N.O.}},
\bauthor{\bsnm{{Meister}}, \binits{C.-V.}}:
\byear{1998},
\batitle{{Large-Scale Electric Fields in Solar Flare Regions}}.
\bjtitle{\solphys}
\bvolume{178},
\bfpage{125}\,--\,\blpage{136}.
\end{barticle}
\endbibitem

\bibitem[\protect\citeauthoryear{{Scherrer}
  \textit{et~al.}}{1995}]{1995SoPh..162..129S}
\begin{barticle}
\bauthor{\bsnm{{Scherrer}}, \binits{P.H.}},
\bauthor{\bsnm{{Bogart}}, \binits{R.S.}},
\bauthor{\bsnm{{Bush}}, \binits{R.I.}},
\bauthor{\bsnm{{Hoeksema}}, \binits{J.T.}},
\bauthor{\bsnm{{Kosovichev}}, \binits{A.G.}},
\bauthor{\bsnm{{Schou}}, \binits{J.}},
\bauthor{\bsnm{{Rosenberg}}, \binits{W.}},
\bauthor{\bsnm{{Springer}}, \binits{L.}},
\bauthor{\bsnm{{Tarbell}}, \binits{T.D.}},
\bauthor{\bsnm{{Title}}, \binits{A.}},
\bauthor{\bsnm{{Wolfson}}, \binits{C.J.}},
\bauthor{\bsnm{{Zayer}}, \binits{I.}},
\bauthor{\bsnm{{MDI Engineering Team}}}:
\byear{1995},
\batitle{{The Solar Oscillations Investigation - Michelson Doppler Imager}}.
\bjtitle{\solphys}
\bvolume{162},
\bfpage{129}\,--\,\blpage{188}.
doi:\doiurl{10.1007/BF00733429}.
\end{barticle}
\endbibitem

\bibitem[\protect\citeauthoryear{{Schindler}, {Hesse}, and
  {Birn}}{1988}]{1988JGR....93.5547S}
\begin{barticle}
\bauthor{\bsnm{{Schindler}}, \binits{K.}},
\bauthor{\bsnm{{Hesse}}, \binits{M.}},
\bauthor{\bsnm{{Birn}}, \binits{J.}}:
\byear{1988},
\batitle{{General magnetic reconnection, parallel electric fields, and
  helicity}}.
\bjtitle{\jgr}
\bvolume{93},
\bfpage{5547}\,--\,\blpage{5557}.
doi:\doiurl{10.1029/JA093iA06p05547}.
\end{barticle}
\endbibitem

\end{thebibliography}
\end{article}
\end{document}